\documentclass[aps,preprint,tightenlines]{revtex4}
\usepackage{graphicx} 
\newcommand{\sh} {/ \hskip-5pt }
\begin{document}
\preprint{\rm FIU-NUPAR-\today{}\\}

\medskip
\medskip

\title{Hard Break-Up of Two-Nucleons from the {\boldmath $^3He$} Nucleus} 
\author{Misak~M.~Sargsian and Carlos Granados}

\affiliation{Florida International University, Miami, Florida 33199, USA\\ \\ \\ }

%\date{\today\\}

\medskip
\medskip

\begin{abstract}
We investigate a large angle photodisintegration of 
two nucleons from the $^3$He nucleus within the framework of 
the hard rescattering model (HRM). In the HRM a  quark of one nucleon 
knocked out by an incoming photon rescatters with a quark of the 
other nucleon leading to the production of two nucleons with large
relative momentum.  Assuming the dominance of the quark-interchange mechanism 
in a hard nucleon-nucleon scattering, the HRM allows the expression of the amplitude of a  two-nucleon break-up reaction through the convolution of photon-quark 
scattering, $NN$ hard scattering amplitude and  nuclear spectral function 
which can be calculated using a nonrelativistic  $^3$He wave function.  
The photon-quark scattering amplitude can be explicitly calculated in the high energy regime, 
whereas for $NN$ scattering one uses the fit of the available experimental data.  
The HRM predicts several specific features for the hard breakup reaction. First, the 
cross section will approximately scale as $s^{-11}$. Secondly, the $s^{11}$ weighted 
cross section will have the shape of energy dependence similar to that of $s^{10}$ weighted $NN$ 
elastic scattering cross section. 
Also one predicts an enhancement of the $pp$ breakup relative to  
the $pn$ breakup cross section as compared to the results from low energy kinematics.
Another result is the prediction of different spectator momentum dependencies of $pp$ 
and $pn$ breakup cross sections. This is due to the fact that same-helicity $pp$-component 
is strongly suppressed in the ground state wave function of $^3$He. Because of this suppression 
the HRM predicts significantly different asymmetries for the cross section of polarization 
transfer $NN$ breakup  reactions for circularly polarized photons.  For the $pp$ breakup this 
asymmetry is predicted to be zero while for the $pn$ it is close to ${2\over 3}$.

\end{abstract}

\maketitle
\section{Introduction}
\label{I}
Two-body breakup processes involving nuclei at high momentum and energy transfer 
play an important role in studies of nuclear QCD. The uniqueness of these processes 
is in the effectiveness by which large values of  invariant 
energy are  produced at rather moderate values of beam energy. The 
kinematics of two-body photodisintegration provide the relation 
\begin{equation}
s_{\gamma NN} \approx   4m_N^2 + 4 E_\gamma m_N,
\label{s}
\end{equation}  
in which the produced invariant energy grows with the energy of the probe  twice as fast as compared, 
for example, to  hard processes involving two protons, in which case $s_{NN} = 2m_N^2 + 2E\cdot m_N$.
As it follows from Eq.(\ref{s}), already at photon energies of $2$~GeV the produced invariant 
mass on one nucleon, $M\sim {\sqrt{S_{\gamma NN}\over 2}}$, exceeds the threshold at which 
deep-inelastic processes become important, $M\gtrsim 2$~GeV.

Combining the above property with a requirement that the momentum 
transfer in the reaction exceeds the masses of the particles involved in the scattering 
($-t,-u\gg m_N^2$) one sets into a hard scattering kinematic regime, in which case  
we expect that  only the minimal Fock components dominate in the wave function of the particles 
involved in the scattering.  Assuming that all the constituents of minimal Fock component participate 
in a hard scattering one arrives at the constituent-counting rule\cite{BF,MMT}. According to this 
rule  we are able to predict the energy dependence of any hard processes 
at fixed and large center of mass~(c.m.) scattering angles. For example, for  
the $A+B\rightarrow C+D$ hard scattering process the constituent-counting rule predicts the 
following energy dependence for the scattering amplitude:
\begin{equation}
{\cal M}\approx s^{-\frac{n_A+n_B+n_C+n_D-4}{2}},
\label{qqrule_M}
\end{equation}
with 
\begin{equation}
\frac{d\sigma}{dt} \sim {|{\cal M}|^2\over s^2}.
\label{crs_scale}
\end{equation}
In Eq.(\ref{qqrule_M}) $n_i$ is the number of constituents in the  minimal Fock component of the 
wave function of particle $i$ involved in the scattering. 
These predictions were confirmed for  a wide 
variety of hard processes involving leptons and hadrons.   

One of the most interesting aspects of the constituent-counting rule is that its application allows 
us to check the onset of quark degrees of freedom in hard reactions involving nuclei~\cite{BCh,Holt}.
This is essential for probing the quark-gluon structure of nuclei.
For example if quarks are involved in hard photodisintegration of the deuteron then according 
to Eqs.(\ref{qqrule_M}) and (\ref{crs_scale}) one expects that 
$\frac {d\sigma}{dt}\sim s^{-11}$~\cite{BCh}.

During the last decade there were several experiments in which 90$^0$ c.m.  photodisintegration of 
the deuteron had been studied at high photon energies 
\cite{NE8,NE17,E89012,Schulte1,gdpnpolexp1,Schulte2,Mirazita,gdpnpolexp2}.
These experiments clearly demonstrated the onset of $s^{-11}$ scaling for the differential 
cross section at 90$^0$ c.m., starting at $E_{\gamma}\ge 1$~GeV. Also, the polarization 
measurements\cite{gdpnpolexp1,gdpnpolexp2} were generally in agreement with the prediction of 
the helicity conservation -- a precursor of the dominance of the mechanism of hard gluon exchange 
involving quarks.

Even though two-body scattering experiments demonstrate clearly an onset of quark degrees of 
freedom in the reaction, they do not affirm the onset of the perturbative QCD (pQCD) regime.  
Indeed it has been argued   that  the validity of constituent-counting rule does not 
necessarily lead to the validity of pQCD(see, e.g., Refs\cite{Isgur_Smith,Rady}). In several measurements
in which the constituent quark rule works pQCD still underestimates  the observed 
cross sections sometimes by several orders of magnitude (see e.g. Refs.\cite{Farrar,BDixon}).  
The latter may indicate a substantial contribution due to nonperturbative effects although
one still may expect sizable contributions from pQCD due to generally unaccounted 
hidden color components in the hadronic and nuclear wave functions\cite{HColor}.

A similar situation also exists for the case of hard photodisintegration of the deuteron. 
Even though experiments clearly indicate the onset of $s^{-11}$ scaling for the cross section of these 
reactions at 90$^0$c.m., one still expects sizable nonperturbative effects. 
Theoretical methods of calculation of these effects are very restricted.  They  
use different approaches to incorporate nonperturbative contributions in the process of 
hard photodisintegration of the deuteron.  The reduced nuclear amplitude~(RNA) formalism  
includes some of the nonperturbative effects through the nucleon form factors\cite{RNA1,RNA2}, 
while
in the quark-gluon string model~(QGS)\cite{QGS} nonperturbative effects are accounted for through the 
reggeization of scattering amplitudes. Also recently, large c.m. angle photodisintegration of the 
deuteron for photon energies up to 2~GeV was calculated within point-form relativistic quantum 
mechanics approximation\cite{MP} in which the strength of the reaction was determined by  short 
range properties of the $NN$ interaction potential.

In the QCD hard rescattering model~(HRM)\cite{gdpn,gdpnpc} it is assumed that the energetic 
photon knocks-out a quark from one nucleon in the deuteron which subsequently experiences hard 
rescattering with a quark of the second nucleon. The latter leads to the production of 
two nucleons with large relative momentum. The summation of all the relevant rescattering diagrams results 
in a scattering amplitude in that the hard rescattering is determined by the large-momentum transfer 
$pn$ scattering amplitude, which includes noncalculable nonperturbative contributions.  
Experimental data are used to estimate the  hard $pn$ scattering amplitude.  The HRM  
allows us to calculate the absolute cross section of   $90^0$ c.m.
hard photodisintegration of the deuteron  without using additional adjustable parameters.

Also,  within the QGS\cite{QGSpol} approximation and the HRM\cite{gdpnpol,hrmictp} 
rather reasonable agreement  has been  obtained for polarization 
observables\cite{gdpnpolexp2}.

Although all the above-mentioned models describe the major characteristics of hard 
photodisintegration  of the deuteron
they are based on very different approaches in the calculation of the nonperturbative 
parts of the photodisintegration reaction.  To investigate further the validity of these 
approaches it was suggested in  Ref.\cite{gheppn} to extend the studies of high energy two-body 
photodisintegration to the case of large angle c.m. breakup of two protons from the $^3$He target. 
In this case not only do the predictions of the above-described models (RNA, QGS, HRM) for 
absolute cross section diverge significantly, but also the  
two-proton breakup reaction from $^3$He  provides additional observables such as 
spectator-neutron momentum distributions that can be used to check further the 
validity of the models. The analysis of the first experimental data on 
$pp$  photodisintegration  of $^3$He nucleus at  high momentum transfer 
is currently in progress at Jefferson Lab\cite{gheppn_prop}. It may significantly advance   
our  understanding of the mechanism of hard breakup reactions involving nuclei.

\medskip
\medskip

In the present work we carry out detailed  studies of reactions involving  hard breakup of 
both $pp$ and $pn$ pairs from the $^3$He target.  We demonstrate that comparative study of 
$pp$ and $pn$ 
breakup processes allows us to gain  new insight into the nature of large c.m. angle scattering. 
One of the observations is that the relative 
strength of $pp$ to $pn$ breakup is 
larger than the one observed in low energy reactions. This is related to 
the onset of quark degrees of freedom in hard breakup reactions in which effectively more 
charges are exchanged between two protons than between proton and neutron. 

Another signature of the HRM is that the shapes of the  energy dependencies of $s^{11}$-scaled differential cross sections 
of $pp$ and $pn$ breakup reactions  mirror the shapes of the energy dependencies of $s^{10}$-scaled differential 
cross sections of hard elastic $pp$ and $pn$  scatterings. 

Within the HRM  one observes also that $pp$ and $pn$ hard breakup processes are sensitive to different 
components of the $^3$He ground state wave function.  This results in different spectator-nucleon 
momentum dependencies for $pp$ and $pn$ hard breakup cross sections.

Because of the different ground state wave function components involved in $pp$ and $pn$ breakup reactions,
the HRM also predicts significantly different magnitudes for  transferred longitudinal polarizations for these 
two processes.

The article is organized as follows: In Sec.~\ref{II}, within the HRM,  we present a detailed 
derivation of the differential cross section of   the reaction of
hard breakup of two-nucleons from a $^3$He target.  In Sec.~\ref{III}  we apply
the formulas derived in the previous section to calculate  the differential cross section of a 
proton-neutron breakup reaction, while in Section~\ref{IV} calculations are done for a 
two-proton breakup reaction. Sec.~\ref{V} considers 
the relative contribution of two- and three-body processes for hard breakup reactions involving 
$A\ge 3$ nuclei. In Sec.~\ref{VI} we present numerical estimates 
for  differential cross sections of $pn$ and $pp$ breakup reactions. 
In Sec.~\ref{VII} we discuss the polarization transfer mechanism of the HRM and estimate the 
asymmetry of the cross section with respect to the helicity of the 
outgoing proton. Results are summarized in Sec.~\ref{VIII}.  
In Appendix~A the details of the derivation of the 
hard rescattering amplitude are given. In Appendix~B we discuss the quark-interchange 
contribution to hard $NN$ elastic scattering.  Appendix~C discusses the method of calculation of 
quark-charge factors within quark-interchange mechanism of $NN$ hard elastic scattering.
In Appendix~D we present the complete list of 
the HRM helicity amplitudes for high energy two-nucleon breakup of the  $^3$He nucleus.

\section{Hard Photodisintegration of Two Nucleons from $^3$He}
\label{II}

\subsection{ Reference frame and kinematics}
\label{RFrame}

\begin{figure}[t]
\centering\includegraphics[height=8.6cm,width=8.6cm]{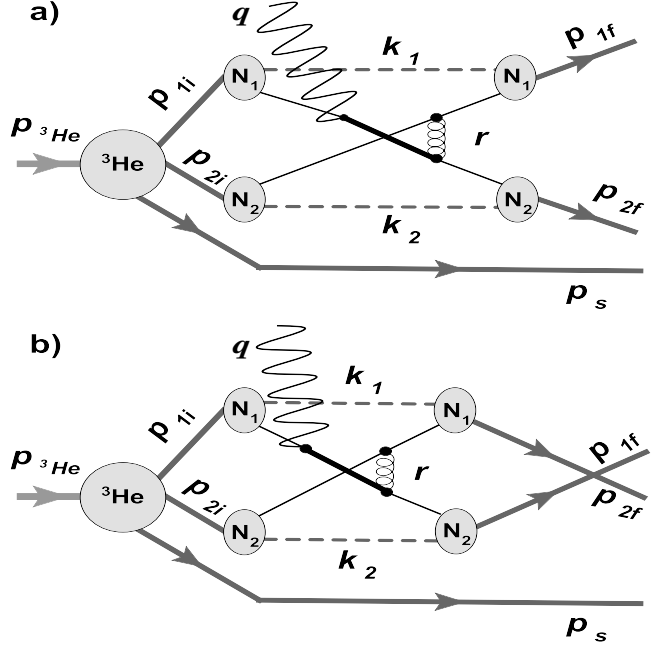}
\caption{(Color online) Typical hard rescattering diagram for the  $NN$ photodisintegration from the $^3$He target.}
\label{Fig.1}
\end{figure}

We are considering a hard photodisintegration of two nucleons  from the $^3$He  
target through the reaction:
\begin{equation}
\gamma + ^3\textnormal{He} \rightarrow (NN) + N_s,
\label{gheNN_N}
\label{reaction}
\end{equation} 
in which two nucleons $(NN)$ are produced at large angles in the ``$\gamma$-$NN$'' center of 
mass reference frame with momenta comparable to the momentum of the initial photon, 
$q$~($>$$1$~GeV/$c$). 
The third nucleon, $N_s$, is produced  with very small momentum $p_s\ll m_N$. 
(Definitions of four-momenta involved in the reaction are given in Fig.\ref{Fig.1}.)

We consider  ``$\gamma$-$NN$'' in a ``$q_+=0$'' reference frame, where 
the light-cone momenta\footnote{The light-cone four-momenta are 
defined as $(p_+,p_-,p_\perp)$, where $p_{\pm} = E\pm p_z$. Here the $z$ axis is defined 
in the direction opposite to the incoming photon momentum.}  of the photon and the $NN$ pair
are defined as follows:
\begin{eqnarray}
 && q^\mu \equiv (q_+,q_-,q_\perp) =   (0, \ \sqrt{s_{NN}'},\ 0), \nonumber \\
 && p_{NN}^\mu \equiv (p_{NN+},p_{NN-},p_{NN\perp}) = (\sqrt{s_{NN}^\prime}, \ 
{M_{NN}^2\over \sqrt{s_{NN}^\prime}}, \ 0),
\label{q+=0}
\end{eqnarray}
where $p_{NN}^\mu=p_{^3\textnormal{He}}^\mu-p_s^\mu$, $M_{NN}^2 = p_{NN}^\mu p_{NN,\mu}$, 
and   $s_{NN}^\prime = s_{NN} - M_{NN}^2$. Here the invariants, $s_{NN}$ and 
$t_{NN}$ are defined as follows:
\begin{eqnarray}
s_{NN} = (q+p_{NN})^2 = (p_{f1}+p_{f2})^2 \nonumber \\ 
t_{NN} = (q - p_{f1})^2 = (p_{f2}-p_{NN})^2.
\label{mans_st}
\end{eqnarray}

As it follows from Eq.(\ref{q+=0}) in the limit of ${M_{NN}^2\over s_{NN}^\prime}\rightarrow 0$ the 
``$q_{+}=0$'' reference frame coincides with  the center of mass frame of the $\gamma$-$NN$ system.
As such it is maximally close to the 
reference frame used  for the $\gamma d\rightarrow p+n$ reaction in Refs.\cite{gdpn} and \cite{gdpnpol}.

\subsection{Hard rescattering model}
\label{Sec.2B}

The hard rescattering model is based on the assumption that in the hard two-nucleon photodisintegration 
reaction, two nucleons with large relative momenta are produced  because the hard rescattering of a fast quark from one nucleon with a quark from the other nucleon.
In this scenario the fast quark is knocked out from a low-momentum nucleon in the nucleus
by an incoming  photon.  
This approach is an alternative to the models in which it is assumed that the incoming 
photon breaks the preexisting  two-nucleon state which has very large relative momentum in 
the nucleus.

The validity of the HRM is based on the observation that the ground state wave functions of light nuclei 
peak strongly at small momenta of bound nucleons, $p\sim 0$. Thus,  diagrams in which an energetic 
photon interacts
with  bound nucleons of small momenta will strongly dominate the diagrams in which  the photon 
interacts with bound  nucleons that have  relative momenta $p\ge  m_N$.

The resulting scenario that the HRM sketches out is as follows (see e.g. Fig.\ref{Fig.1}): first,the  
incoming photon will  knock out 
a quark from one of the nucleons in the nucleus and then the struck quark that now carries almost 
the whole
momentum of the photon  will share its momentum 
with a quark from the other nucleon  through the exchanged gluon.  The resulting two energetic 
quarks will recombine with the residual quark-gluon systems to produce two nucleons with large relative 
momentum ~($\sim$$q$). This recombination will contain  gluon exchanges and also 
incalculable nonperturbative interactions.

Note that for the quark-gluon picture discussed above to be relevant the intermediate masses 
$m_{\textnormal{\scriptsize int}}$ produced after 
the photon absorption should exceed the  mass scale  characteristic for deep inelastic scattering, 
$W\sim 2.2$~GeV.
Using the relation $m_{\textnormal{\scriptsize int}}\approx \sqrt{m_N^2 + 2E_{\gamma}m_N}$, from the requirement that 
$m_{\textnormal{\scriptsize int}}\ge W$ 
one obtains the condition $E_\gamma\ge 2$~GeV. Additionally, to ensure the validity of quark degrees 
of freedom in the final state rescattering, one requires  $k_{\textnormal{\scriptsize rel}}\ge 1$~GeV/$c$
for the relative momentum $k_{\textnormal{\scriptsize rel}}$, of two outgoing 
nucleons. All of these imposes a restriction on the incoming photon energy,
$E_\gamma \ge 2$~GeV, and for transferred momenta $-t,-u\ge 2$~GeV$^2$. 
Note that provided a smooth transition from hadronic to quark-gluon degrees of freedom 
in nuclei one expects the validity of the HRM to extend to even lower values of $E_\gamma$~
($\gtrsim$$1$~GeV). This expectation was confirmed in recent measurements of angular dependencies 
of the $\gamma d\rightarrow pn$ cross section for a wide range of incoming photon energies\cite{Mirazita}.

\medskip
\medskip

To calculate the differential cross section of the hard photodisintegration reaction of 
Eq.(\ref{gheNN_N}) within  the HRM one needs to evaluate the sum of   hard rescattering diagrams 
similar to the one presented in Fig.\ref{Fig.1}.
We start with analyzing the scattering amplitude corresponding to the 
diagrams of Fig.\ref{Fig.1}. Using Feynman rules and applying the light-cone 
wave function reduction described in Appendix~A,  we obtain
\begin{eqnarray}
& & \langle\lambda_{1f},\lambda_{2f},\lambda_s\mid A \mid \lambda_\gamma, \lambda_A\rangle = 
\sum\limits_{(\eta_{1f},\eta_{2f}),(\eta_{1i},\eta_{2i}),(\lambda_{1i},\lambda_{2i})} \int 
\left\{ {\psi^{\dagger\lambda_{2f},\eta_{2f}}_N(p_{2f},x'_2,k_{2\perp})\over 1-x'_2}\bar 
u_{\eta_{2f}}(p_{2f}-k_2) \right. 
\nonumber \\ & & 
[-igT^F_c\gamma^\nu]
{i[\sh p_{1i}-\sh k_1 + \sh q + m_q]\over (p_{1i}-k_1+q)^2-m_q^2 + i\epsilon}
[-iQ_ie{\bf \epsilon^{\lambda_\gamma}_\perp\gamma^\perp}]u_{\eta_{1i}}(p_{1i}-k_1)
\left. {\psi_N^{\lambda_{1i},\eta_{1i}}(p_{1i},x_1,k_{1\perp})\over (1-x_1)}\right\}_1\times
\nonumber \\ & & 
\left\{ 
{\psi^{\dagger\lambda_{1f},\eta_{1f}}_N(p_{1f},x'_1,k_{1\perp})\over 1-x'_1}
\right.  
 \bar u_{\eta_{1f}}(p_{1f}-k_1)[-igT^F_c\gamma^\mu]u_{\eta_{2i}}(p_{2i}-k_2) 
\left . {\psi_N^{\lambda_{2i},\eta_{2i}}(p_{2i},x_2,k_{2\perp})\over (1-x_2)} \right\}_2\times
\nonumber \\ & &
G^{\mu,\nu}(r) {dx_1\over x_1}{d^2k_{1\perp}\over 2 (2\pi)^3}
{dx_2\over x_2}{d^2k_{2\perp}\over 2 (2\pi)^3} 
{\Psi_{^3\textnormal{\scriptsize {He}}}^{\lambda_A,\lambda_{1i},\lambda_{2i},\lambda_s}(\alpha,{p_\perp},p_s)
\over (1-\alpha)}{d\alpha\over \alpha}
{d^2p_{\perp}\over 2(2\pi)^3}  -  \left(p_{1f}\longleftrightarrow p_{2f}\right),
\label{ampl0}
\end{eqnarray}
where the $\left(p_{1f}\longleftrightarrow p_{2f}\right)$ part accounts for the diagram in 
Fig.\ref{Fig.1}(b).
Here the four-momenta, $p_{1i}$, $p_{2i}$, $p_s$, $k_1$, $k_2$, $r$, $p_{1f}$ and $p_{2f}$ 
are defined in Fig.\ref{Fig.1}. Note that $k_1$ and $k_2$ define
the four-momenta of residual quark-gluon system of the nucleons without specifying 
their actual composition. We also define  $x_1$, $x'_1$, 
$x_2$ and $x'_2$ as the light-cone  momentum fractions of initial and final nucleons 
carried by their respective residual quark-gluon systems: $x_{1(2)} = {k_{1(2)+}\over p_{1(2)i+}}$ 
and $x'_{1(2)} = {k_{1(2)+}\over p_{1(2)f+}}$. For the $^3$He wave function,
$\alpha={p_{2+}\over p_{NN+}}$ is 
the light-cone momentum fraction of the $NN$ pair carried by one of the nucleons in the pair, and 
$p_{\perp}$ is their relative transverse momentum.
The scattering process in Eq.(\ref{ampl0}) can be described through the combination 
of the following blocks: 
(a)~$\Psi_{^3\textnormal{\scriptsize {He}}}^{\lambda_A,\lambda_{1i},\lambda_{2i},\lambda_s}(\alpha,{p_{\perp}},p_s)$, 
is the light-cone $^3$He-wave function that describes a transition of the $^3$He nucleus with 
helicity $\lambda_A$ into three nucleons with $\lambda_{1i}$ , $\lambda_{2i}$, and $\lambda_s$ helicities, 
respectively. 
(b)~ The term in $\{...\}_{1}$ describes the ``knocking out'' of 
an $\eta_{1i}$-helicity quark  from a $\lambda_{1i}$-helicity nucleon by an incoming 
photon with helicity $\lambda_\gamma$.
Subsequently, the knocked-out  quark exchanges a gluon,  
($[-igT^F_c\gamma^\nu]$), with a quark from the second nucleon producing a final 
$\eta_{2f}$-helicity quark that combines into the nucleon ``${2f}$'' with helicity $\lambda_{2f}$. 
(c)~The term in $\{...\}_{2}$ describes the emerging $\eta_{2i}$-helicity
quark from the $\lambda_{2i}$ -helicity nucleon
then  then exchanges a gluon, ($[-igT^F_c\gamma^\mu]$), with the knocked-out 
quark and produces a final $\eta_{1f}$-helicity quark that combines into the  
nucleon ``${1f}$'' with helicity $\lambda_{1f}$.  
d)~The propagator of the exchanged gluon is 
$G^{\mu\nu}(r) = {d^{\mu\nu}\over r^2+i\varepsilon}$ with polarization matrix, 
$d^{\mu\nu}$ (fixed by the light-cone gauge), and $r=(p_2-k_2+l)-(p_1-k_1+q)$, with 
$l = (p_{2f}-p_{2i})$. In Eq.(\ref{ampl0}) the $\psi^{\lambda,\eta}_N$ represents everywhere an 
$\eta$-helicity  single quark  wave function of a $\lambda$-helicity nucleon 
as defined in Eq.(\ref{nwf}) and $u_\tau$ is the quark spinor defined in the helicity basis.

We now consider the denominator of the struck quark's propagator which can be  represented as follows:
\begin{equation}
(p_{1i}-k_1 +q)^2 - m_q^2+i\varepsilon = 
(1-x_1)s_{NN}^\prime(\alpha_c - \alpha+i\epsilon),
\label{denom}
\end{equation}
where
\begin{equation}
\alpha_c = 1 + {1\over s_{NN}'}\left[\tilde m_N^2 - 
{ m_s^2(1-x_1)+m_q^2x_1+(k_1-x_1p_1)^2\over x_1(1-x_1)}\right]
\label{alphac}
\end{equation}
Here $m_s^2$ and $\tilde m_N^2\approx m_N^2$ are defined in Eqs.(\ref{k1onshell}) and (\ref{tilmass}), 
and  $m_q$ represents the current quark mass of the knocked out quark. 
Our further discussion is based on the use of the fact 
that the  $^3$He wave function  strongly peaks at $\alpha={1\over 2}$, which corresponds to the
kinematic situation in which two constituent nucleons have equal share of the $NN$ pair's 
light-cone momentum.  Thus one expects that the integral in Eq.(\ref{ampl0}) is dominated by 
the value of the integrand at $\alpha=\alpha_c = {1\over 2}$.   This allows us to perform 
$\alpha$-integration in Eq.(\ref{ampl0}) through the pole of the denominator (\ref{denom}) 
at $\alpha=\alpha_c$ (i.e. keeping only the $-i\pi \delta(\alpha-\alpha_c)$ part of 
the relation 
${1\over \alpha_c-\alpha+i\epsilon} = -i\pi\delta(\alpha-\alpha_c) + P{1\over \alpha_c-\alpha}$) 
and later replacing $\alpha_c$ by ${1\over 2}$.
Using this relation 
to estimate the propagator of the struck quark at its on-mass shell value ($\alpha=\alpha_c$)  allows us 
to represent,
$(\sh p_{1i} - \sh k_1 + \sh q)^{\textnormal{\scriptsize {on \ shell}}} + m_q = \sum_\zeta u_\zeta 
(p_1-k_1+q)\bar u_\zeta(p_1-k_1+q)$. Then  for the 
scattering amplitude of Eq.(\ref{ampl0}) one obtains
\begin{eqnarray}
& & \langle\lambda_{1f},\lambda_{2f},\lambda_s\mid A_i \mid \lambda_\gamma, \lambda_A\rangle = 
\nonumber \\ & & 
\sum\limits_{(\eta_{1f},\eta_{2f}),(\eta_{1i},\eta_{2i}),(\lambda_{1i},\lambda_{2i}),\zeta} \int 
\left\{ {\psi^{\dagger\lambda_{2f},\eta_{2f}}_N(p_{2f},x'_2,k_{2\perp})\over 1-x'_2}\bar 
u_{\eta_{2f}}(p_{2f}-k_2) [-igT^F_c\gamma^\nu]\right. \nonumber \\ & & 
{i\cdot u_\zeta(p_1-k_1+q)\bar u_\zeta(p_1-k_1+q)\over (1-x_1)s'}
[-iQ_ie{\bf \epsilon^{\lambda_\gamma}_\perp\gamma^\perp}]u_{\eta_{1i}}(p_{1i}-k_1)
\left. {\psi_N^{\lambda_{1i},\eta_{1i}}(p_{1i},x_1,k_{1\perp})\over (1-x_1)}\right\}_1
\nonumber \\ & & 
\left\{ {\psi^{\dagger\lambda_{1f},\eta_{1f}}_N(p_{1f},x'_1,k_{1\perp})\over 1-x'_1}\right.  
 \bar u_{\eta_{1f}}(p_{1f}-k_1)[-igT^F_c\gamma^\mu]u_{\eta_{2i}}(p_{2i}-k_2) 
\left . {\psi_N^{\lambda_{2i},\eta_{2i}}(p_{2i},x_2,k_{2\perp})\over (1-x_2)} \right\}_2 
\nonumber \\ & & 
G^{\mu,\nu}(r) {dx_1\over x_1}{d^2k_{1\perp}\over 2 (2\pi)^3}
{dx_2\over x_2}{d^2k_{2\perp}\over 2 (2\pi)^3} 
{\Psi_{^3\textnormal{\scriptsize {He}}}^{\lambda_A,\lambda_{1i},\lambda_{2i},\lambda_s}(\alpha_c,p_{\perp},p_s)\over (1-\alpha_c) 
\alpha_c}
{d^2p_{\perp}\over 4(2\pi)^2}-     \left(p_{1f}\longleftrightarrow p_{2f}\right).
\label{ampl1}
\end{eqnarray}

Next we evaluate the matrix element of the photon-quark interaction using
on-mass shell spinors for the struck quark.
Taking into account the fact that $(p_{1i}-k)_{+}\gg |k_{\perp}|, m_q$, for this matrix element 
we obtain
\begin{eqnarray}
\bar u_\zeta(p_{1i} - k_{1} + q)[-iQ_ie \epsilon^{\lambda_\gamma}_\perp \gamma^\perp]u_{\eta_{1i}}(p_{1i}-k_1) = 
ieQ_i2\sqrt{2 E_2E_1}(\lambda_\gamma)\delta^{\lambda_\gamma\zeta}\delta^{\lambda_\gamma\eta_{1i}}
\label{melement}
\end{eqnarray}
where $E_{1}= (1-\alpha)(1-x_1){\sqrt{s_{NN}^\prime}\over 2}$ and $E_2 = (1-(1-\alpha)(1-x_1)){\sqrt{s_{NN}^\prime}\over 2}$.

Further explicit calculations of Eq.(\ref{ampl1}) require the knowledge of  quark wave functions 
of the nucleon. Also, one needs to sum over the multitude of the amplitudes representing 
different topologies of  quark knock-out  rescattering and recombinations into 
two final nucleon states.

We attempt to solve this problem by using the above observation that 
the $^3$He wave function strongly peaks at $\alpha = {1\over 2}$.
We  evaluate Eq.(\ref{ampl1}) setting everywhere  $\alpha_c = {1\over 2}$. 
Such approximation significantly simplifies further derivations. 
As it follows from Eq.(\ref{alphac}) the $\alpha_c={1\over 2}$ condition restricts the
values of $x_1$ of the recoil quark-gluon system to  $x_1 \sim {k_{1\perp}^2\over s_{NN}^\prime}$, thereby 
ensuring that the quark-interchange happens for the valence quarks with $x_q= 1-x_1\sim 1$.
The latter allows us to simplify Eq.(\ref{melement}) setting $E_{1}=E_{2} = {\sqrt{s^\prime_{NN}}\over 4}$.
Using these approximations and substituting Eq.(\ref{melement}) into Eq.(\ref{ampl1}) one obtains
\begin{eqnarray}
& & \langle\lambda_{1f},\lambda_{2f}\lambda_s\mid A_i \mid \lambda_\gamma, \lambda_A\rangle = 
i[\lambda_\gamma]e\sum\limits_{(\eta_{1f},\eta_{2f}),(\eta_{2i}),(\lambda_{1i},\lambda_{2i})} \int 
{Q_i \over \sqrt{2s'}}
\times  \nonumber \\& &
\left[ \left\{ {\psi^{\dagger\lambda_{2f},\eta_{2f}}_N(p_{2f},x'_2,k_{2\perp})\over 1-x'_2}\bar 
u_{\eta_{2f}}(p_{2f}-k_2) [-igT^F_c\gamma^\nu]\right. \right. 
u_{\lambda_\gamma}(p_1-k_1+q)
\left. 
{\psi_N^{\lambda_{1i},\lambda_\gamma}(p_{1i},x_1,k_{1\perp})\over (1-x_1)} 
\right\}  
\nonumber \\
& & \left\{ 
{\psi^{\dagger\lambda_{1f},\eta_{1f}}_N(p_{1f},x'_1,k_{1\perp})\over 1-x'_1}
\bar u_{\eta_{1f}}(p_{1f}-k_1)[-igT^F_c\gamma^\mu] \right.  
\left. u_{\eta_{2i}}(p_{2i}-k_2) 
 {\psi_N^{\lambda_{2i},\eta_{2i}}(p_{2i},x_2,k_2)\over (1-x_2)} \right\} \nonumber \\
& & \left. G^{\mu,\nu}(r) {dx_1\over x_1}{d^2k_{1\perp}\over 2 (2\pi)^3}
{dx_2\over x_2}{d^2k_{2\perp}\over 2 (2\pi)^3} \right]_{\textnormal{\scriptsize {QIM}}}
\Psi_{^3\textnormal{\scriptsize {He}}}^{\lambda_A,\lambda_{1i},\lambda_{2i}}(\alpha={1\over 2},p_{2\perp)}
{d^2p_{2\perp}\over (2\pi)^2} \ \ \ \ \   - \left(p_{1f}\leftrightarrow p_{2f}\right).
\label{ampl2}
\end{eqnarray}
Note that due to the $\delta$ factors in Eq.(\ref{melement}) the helicity of the knocked out 
quark in Eq.(\ref{ampl2}) is equal to the helicity of incoming photon, that is 
$\eta_{1i}=\lambda_{\gamma}$.

To proceed, we observe that the kernel, $[\dots ]_{\textnormal{\scriptsize QIM}}$  representing the quark-interchange 
mechanism (QIM) of the rescattering in  Eq.(\ref{ampl2}) can 
be identified with the quark-interchange contribution in the $NN$ scattering amplitude (see Appendix B).
Such identification can be done by observing that in the chosen reference frame, $q_+=0$, and 
the quark wave function of the nucleon depends on the quark's light-cone momentum fraction and  transverse 
momentum only, which are the same in both Eqs.(\ref{ampl2}) and (\ref{NN_qim}).
For our derivation we also use the above-discussed observation that the  
$\alpha=\alpha_c={1\over 2}$ condition ensures that 
the quark-interchange happens for the valence quarks with $x_q= 1-x_1\sim 1$. This justifies
our next assumption, that valence quarks carry  the helicity of their parent nucleon 
(i.e. $\eta_{1i} = \lambda_i$).   The last assumption allows us to perform the summation of 
Eq.(\ref{ampl2}) over the helicities of the exchanged quarks ($\eta_{2i},\eta_{1f},\eta_{2f}$)  
and to use Eq.(\ref{NN_qim}) to express  
the QIM part   in Eq.(\ref{ampl2}) through the corresponding QIM amplitude of $NN$ scattering. 
Summing for all possible topologies of quark-interchange  diagrams we arrive at
\begin{eqnarray}
& & \langle\lambda_{1f},\lambda_{2f},\lambda_s\mid A \mid \lambda_\gamma, \lambda_A\rangle 
= ie[\lambda_\gamma]\times  \nonumber \\
& & \left\{ \sum\limits_{i \in N_1}\sum\limits_{\lambda_{2i}} \int 
{Q_i^{N_1}\over \sqrt{2s'}}
\langle \lambda_{2f};\lambda_{1f}\mid T_{NN,i}^{\textnormal{\scriptsize {QIM}}}(s,l^2)\mid \lambda_\gamma;\lambda_{2i}\rangle 
\Psi_{^3\textnormal{\scriptsize {He}}}^{\lambda_A}(p_1,\lambda_\gamma;p_2,\lambda_{2i},p_s,\lambda_s)
{d^2p_\perp \over (2\pi)^2} \right. \nonumber \\ 
& + &\left.
\sum\limits_{i \in N_2}\sum\limits_{\lambda_{1i}} \int 
{Q_i^{N_2}\over \sqrt{2s'}}
\langle \lambda_{2f};\lambda_{1f}\mid T_{NN,i}^{\textnormal{\scriptsize {QIM}}}(s,l^2)\mid \lambda_{1i};\lambda_{\gamma}\rangle 
\Psi_{^3\textnormal{\scriptsize {He}}}^{\lambda_A}(p_1,\lambda_{1i};p_2,\lambda_{\gamma},p_s,\lambda_s)
{d^2p_\perp \over (2\pi)^2} \right\}
\label{ampl4}
\end{eqnarray}
where nucleon momenta $p_1$ and $p_2$ have half of their c.m.  momentum fractions  and $p_\perp$ is their 
relative transverse momentum with respect to the direction of the photon momentum [see Eq.(\ref{momrel})].
Here, for example,  $Q_i^{N}\cdot \langle \lambda_{2f};\lambda_{1f}\mid T_{NN,i}^{\textnormal{\scriptsize {QIM}}}(s,l^2)\mid 
\lambda_1;\lambda_{2}\rangle$ 
represents the quark-interchange amplitude of $NN$ interaction weighted with the charge of those  
interchanging quarks $Q^{N}_i$ that are struck from a nucleon $N$ by the incoming photon.
The sum ($\sum\limits_{i \in N}$) can be performed within the quark-interchange model of  $NN$ interaction 
which allows us to represent the $NN$ scattering amplitude as follows~\cite{BCL}:
\begin{equation} 
\langle a'b' |T_{NN}^{\textnormal{\scriptsize {QIM}}}|ab\rangle = 
{1\over 2}\langle a'b'| \sum\limits_{i\in a\ , \ j\in b} 
[I_iI_j + \vec \tau_i\vec\cdot\tau_j] F_{i,j}(s,t)|ab\rangle
\label{NNQIM}
\end{equation}
where $I_i$ and  $\tau_i$ are the identity and  Pauli matrices 
defined in the SU(2) flavor (isospin) space of the interchanged 
quarks. The kernel $F_{i,j}(s,t)$ describes an interchange of $i$ and $j$ 
quarks.~\footnote{The additional assumption of helicity conservation allows us to 
express the kernel in the form\cite{BCL} 
$F_{i,j}(s,t) = {1\over 2}[I_iI_j + \vec \sigma_i\vec\cdot\sigma_j]\tilde F_{i,j}(s,t)$, 
where $I_i$ and  $\sigma_i$ operate in the SU(2) helicity ($H$-spin) space of 
exchanged ($i,j$) quarks\cite{BCL}. However for our discussion the assumption of 
helicity conservation is not required.}.

Using Eq.(\ref{NNQIM}) one can calculate the quark-charge weighted QIM 
amplitude, $Q_i\cdot \langle a'b' |T_{NN,i}^{\textnormal{\scriptsize {QIM}}}|ab\rangle$, as follows:
\begin{eqnarray}
 \sum\limits_{i \in N} Q^{N}_i\langle a'b' |T_{NN,i}^{\textnormal{\scriptsize {QIM}}}|ab\rangle   &  = & 
{1\over 2}\langle a'b'| \sum\limits_{i\in a\ , \ j\in b} 
[I_iI_j + \vec \tau_i\vec\cdot\tau_j] (Q_i)F_{i,j}(s,t)|ab\rangle \nonumber\\
&  = & Q^N_F\cdot \langle a'b' |T_{NN}^{\textnormal{\scriptsize {QIM}}}|ab\rangle,
\label{QNNQIM}
\end{eqnarray}
where $Q^N_F$ are the charge factor that are  explicitly calculated using the method described in 
Appendix C. These factors can be expressed through the combinations of  valence quark charges $Q_i$ of nucleon $N$ and 
the number of quark interchanges for each flavor of quark, $N_{Q_i}$, necessary to 
produce a given helicity $NN$ amplitude,  as follows,
\begin{equation}
Q^{N}_F= {N_{uu}(Q_u) + N_{dd}(Q_d) + N_{ud}(Q_u+Q_d)\over 
N_{uu} + N_{dd} + N_{ud}}.
\label{QF}
\end{equation}

\medskip
\medskip

Next we discuss the light-cone wave function of $^3$He that enters in Eq.(\ref{ampl4}). 
The important result 
that allows us to evaluate the wave function is the observation that two nucleons that interact with the photon 
share equally the $NN$ pair's c.m. momentum ($p_{NN}$), i.e., $\alpha = {1\over 2}$.  
If we constrain the third nucleon's light-cone momentum fraction 
$\alpha_s  = {3\cdot p_{s+}\over p_{^3\textnormal{\scriptsize He+}}} = {3(E_s + p_s^z)\over E_{^3\textnormal{\scriptsize {He}}}+ p_s^z + p_{NN}^z} \approx 1$ 
and  transverse momentum $p_{s\perp}\ll m_N$, then the momenta of all the nucleons in the nucleus 
are nonrelativistic. In this case one can use the calculation of triangle diagrams,
which provides the normalization of nuclear wave functions based on baryonic number 
conservation to relate LC and  nonrelativistic nuclear wave functions  as follows\cite{FS81,FS88}
\begin{equation}
\Psi_{^3\textnormal{\scriptsize {He}}}(\alpha,p_\perp, \alpha_s, p_{s,\perp}) =
\sqrt{2}(2\pi)^3m_N \Psi_{^3\textnormal{\scriptsize {He,NR}}}(\alpha,p_{\perp},\alpha_s,p_{s,\perp})
\label{wfnr}
\end{equation}
where for  $\Psi_{^3\textnormal{\scriptsize{He,NR}}}$ we can use known nonrelativistic $^3$He wave functions 
(see e.g.,\cite{Nogga}).

Substituting Eqs.(\ref{QNNQIM}) and (\ref{wfnr}) into Eq.(\ref{ampl4}) for the two-nucleon 
photodisintegration amplitude we obtain
\begin{eqnarray}
& & \langle\lambda_{1f},\lambda_{2f},\lambda_s\mid M \mid \lambda_\gamma, \lambda_A\rangle =  
{i[\lambda_\gamma]e\sqrt{2}(2\pi)^3\over \sqrt{2S^\prime_{NN}}}\times \nonumber \\
& & \ \ \left\{ Q_F^{N_1} \sum\limits_{\lambda_{2i}} \int  
\langle \lambda_{2f};\lambda_{1f}\mid T_{NN}^{\textnormal{\scriptsize {QIM}}}(s_{NN},t_{N})\mid \lambda_\gamma;\lambda_{2i}\rangle 
\Psi_{^3\textnormal{\scriptsize {He,NR}}}^{\lambda_A}
(\vec p_1,\lambda_\gamma;\vec p_2,\lambda_{2i};\vec p_s,\lambda_s)m_N{d^2p_{\perp} \over (2\pi)^2} + \right. \nonumber \\
& & \ \ \left. Q_F^{N_2}  \sum\limits_{\lambda_{1i}} \int  
\langle \lambda_{2f};\lambda_{1f}\mid T_{NN}^{\textnormal{\scriptsize {QIM}}}(s_{NN},t_{N})\mid \lambda_{1i};\lambda_{\lambda}\rangle 
\Psi_{^3\textnormal{\scriptsize {He,NR}}}^{\lambda_A}
(\vec p_1,\lambda_{1i};\vec p_2,\lambda_{\gamma};\vec p_s,\lambda_s)m_N{d^2p_{\perp} \over (2\pi)^2}\right\}
\nonumber \\
\label{ampl5}
\end{eqnarray}
where in the Lab frame of the $^3$He nucleus, defining the $z$ direction along the 
direction of $q_{\textnormal{\scriptsize Lab}}$ one obtains
\begin{eqnarray}
&  \alpha = {E_2 - p_{2z}\over M_A-E_s - p_{sz}}; \ \ \ \  &   
p_{\perp} = {p_{1\perp}-p_{2\perp}\over 2}, \nonumber \\
&  \alpha_s = {E_s + p_{sz}\over M_A/A}; &  \vec p_1+ \vec p_2 = - \vec p_s,
\label{momrel}
\end{eqnarray}
with all the momenta defined in the Lab frame.

Equation(\ref{ampl5}) allows us to calculate the unpolarized differential cross section of two nucleon 
breakup in the form
\begin{equation}
{d\sigma\over dt d^3p_{s}/(2E_s (2\pi)^3)} = {|\bar {\cal M}|^2\over 16\pi (s-M_A^2)(s_{NN}-M^2_{NN})}
\label{hrm_crs}
\end{equation}
where $s = (k_\gamma + p_{A})^2$ and 
\begin{equation}
|\bar M|^2 = {1\over 2}\cdot{1\over 2}\sum\limits_{\lambda_{1f},\lambda_{2f},\lambda_s,\lambda_{\gamma},\lambda_A}
\left|\langle\lambda_{1f},\lambda_{2f},\lambda_s\mid M \mid \lambda_\gamma, \lambda_A\rangle\right|^2.
\label{M2}
\end{equation}

As follows from Eq.(\ref{ampl5}) the knowledge of quark-interchange helicity amplitudes of $NN$ 
elastic scattering will allow us to calculate the differential cross section of hard $NN$ breakup 
reaction without introducing any adjustable  parameter.

\medskip
\medskip

Because the assumption of  $\alpha_c={1\over 2}$ plays a major role in the above derivations 
we attempt now to estimate the theoretical error introduced by this approximation. This approximation
by its nature is a ``peaking'' approximation that is used in loop calculations involving Feynman 
diagrams (one such example is the calculation of radiative effects in electroproduction processes 
;see, e.g., \cite{MoTs}).  One way to estimate the accuracy of the approximation is to identify 
the main dependence  of the integrand in  Eq.(\ref{ampl1}) on $\alpha_c$ which can be evaluated exactly 
and compare with its evaluation at $\alpha_c={1\over 2}$.  Using Eq.(\ref{melement}) as well as 
Eq.(\ref{alphac}) that allows us to  relate ${dx_1\over x_1}$ to ${d\alpha_c\over \alpha_c}$, and assuming 
that the quark wave functions of nucleons at $\alpha_c\sim {1\over 2}$ are less  sensitive to  $\alpha_c$, 
one arrives at
\begin{equation}
R(p_s) =  {4 \Psi_{^3\textnormal{\scriptsize {He}}}^{\lambda_A,\lambda_{1i},\lambda_{2i},\lambda_s}(\alpha_c={1\over 2},p_{\perp},p_s) \over 
\int {d\alpha_c\over \alpha_c} 
{\Psi_{^3\textnormal{\scriptsize {He}}}^{\lambda_A,\lambda_{1i},\lambda_{2i},\lambda_s}(\alpha_c,p_{\perp},p_s)\over 
\sqrt{(1-\alpha_c) \alpha_c}}}.
\end{equation}
This ratio depends on the kinematics of the spectator nucleon, and for the case of $p_s\le 100$~MeV/$c$, 
$R(p_s)\approx 1.1$, which corresponds to $\sim$$20\%$ of uncertainty in the cross section of the reaction
calculated with the $\alpha_c={1\over 2}$ approximation.
The uncertainty increases with an increase of the momentum of the spectator nucleon.
This can be understood qualitatively because, for large center of mass momenta of the $NN$ pair, 
the $\alpha={1\over 2}$ peak of the nuclear  wave function is less pronounced.

\subsection{Quark-interchange and hard $NN$ elastic scattering amplitudes}
The possibility of using $NN$ elastic scattering data to calculate the 
cross section in Eqs.(\ref{hrm_crs}) and (\ref{M2}) is based on the assumption that the quark-interchange 
mechanism provides the bulk of the $NN$ elastic scattering strength at high energies and large c.m. angles.
This is a rather well-justified assumption. Experiments on exclusive large $-t$ two-body 
reactions~\cite{20ht} demonstrated clearly the  dominance of  the quark-interchange mechanism for   
the scattering of  hadrons that share common quark flavors.
The analysis of these  experiments  indicate that  
contributions from competing mechanisms such as pure gluon exchange or quark-antiquark 
annihilation are on the level of few percent. This justifies our next approximation, 
to  substitute quark-interchange $NN$ amplitudes in Eq.(\ref{ampl5}) 
with actual $NN$ helicity amplitudes as follows:
\begin{eqnarray}
<+,+|T^{\textnormal{\scriptsize {QIM}}}_{NN}|+,+>  &=& \phi_1 \nonumber \\
<+,+|T^{\textnormal{\scriptsize {QIM}}}_{NN}|+,->  &=& \phi_5 \nonumber \\
<+,+|T^{\textnormal{\scriptsize {QIM}}}_{NN}|-,->  &=& \phi_2  \nonumber \\
<+,-|T^{\textnormal{\scriptsize {QIM}}}_{NN}|+,->  &=& \phi_3 \nonumber \\
<+-|T^{\textnormal{\scriptsize {QIM}}}_{NN}|-+>  &=&  -\phi_4.
\label{phis}
\end{eqnarray}
All other helicity combinations can be related to the above amplitudes through 
the parity and time-reversal symmetry. The minus sign in the 
last equation above is due to the Jackob-Wick phase factor (see, e.g., Ref.\cite{FGST}), according to 
which  one gains a phase factor of ($-$$1$) if two quarks that scatter
by $\pi-\theta$ angle in c.m. have opposite helicities~\cite{JW}. Note that 
$\phi_i$'s are normalized 
in such a way that the cross section for $NN$ scattering is defined as
\begin{eqnarray}
{d\sigma^{NN\rightarrow NN}\over dt} =  {1\over 16\pi}{1\over s(s-4m_N^2)}
{1\over 2}(|\phi_1|^2 + 
|\phi_2|^2 + |\phi_3|^2 + |\phi_4|^2 + 4|\phi_5|^2).
\label{crs_NN}
\end{eqnarray}

Because in the hard breakup regime the momentum transfer $-t_N\gg m_N^2$, one can factorize  the helicity 
$NN$ amplitudes from Eq.(\ref{ampl5}) at $s_{NN}$ and  $t_{N}$ values  
defined as follows:
\begin{eqnarray}
s_{NN}  & = & (q + p_{NN})^2 = (p_{f1} + p_{f2})^2, \nonumber \\
t_{N} & = & (p_{f2} - p_{NN}/2)^2   = {t_{NN}\over 2} + {m_N^2\over 2}  - {M_{NN}^2\over 4}.
\label{stuN}
\end{eqnarray}
Using this  factorization in Eq.(\ref{ampl5}) for the spin averaged square of the breakup amplitude
one obtains
\begin{eqnarray}
\bar {|{\cal M} |}^2 = & &  {(e^2 2(2\pi)^6\over 2 s^\prime_{NN}}
{1\over 2}\left\{2 Q^2_F|\phi_5|^2 S_0 + Q_F^2(|\phi_1|^2+|\phi_2|^2)S_{12} + \right. \nonumber \\ 
& &\ \ \ \ \ \ \ \ \ \ \ \left. \left[ (Q_F^{N_1}\phi_3 + Q_F^{N_2}\phi_4)^2 +
(Q_F^{N_1}\phi_4+Q_F^{N_2}\phi_3)^2)\right]S_{34}\right\},
\label{M2_fct}
\end{eqnarray}
where $Q_F = Q_{F}^{N_1} + Q_{F}^{N_2}$ and $S_{12}$, $S_{34}$, and $S_0$ are partially integrated nuclear spectral 
functions:
\begin{eqnarray}
S_{12}(t_1,t_2,\alpha,\vec p_s) = N_{NN}\sum\limits_{\lambda_1=\lambda_2=-{1\over2}}^{1\over 2}
\sum\limits_{\lambda_3 
= -{1\over2}}^{1\over 2} \left|\int \Psi_{^3\textnormal{\scriptsize {He,NR}}}^{1\over 2}
(\vec p_1,\lambda_1,t_1;\vec p_2,\lambda_{2},t_2;\vec p_s,\lambda_3)
m_N{d^2p_{\perp} \over (2\pi)^2}\right|^2,
\label{S12}
\end{eqnarray}
\begin{eqnarray}
S_{34}(t_1,t_2,\alpha,\vec p_s) = N_{NN}\sum\limits_{\lambda_1=-\lambda_2=-{1\over2}}^{1\over 2}
\sum\limits_{\lambda_3 = -{1\over2}}^{1\over 2}
\left| \int \Psi_{^3\textnormal{\scriptsize {He,NR}}}^{1\over 2}
(\vec p_1,\lambda_1,t_1;\vec p_2,\lambda_{2},t_2;\vec p_s,\lambda_3)
m_N{d^2p_{\perp} \over (2\pi)^2}\right|^2
\label{S34}
\end{eqnarray}
and
\begin{equation}
S_0 = S_{12}+S_{34}.
\label{S0}
\end{equation}
In the above equations $t_1$ and $t_2$ are the isospin projections of nucleons 
in $^3$He. The wave function is normalized to ${2\over 3}$ for proton and ${1\over 3}$ for neutron.
The normalization constants, $N_{NN}$  renormalize the wave function to one $pp$ and two $np$ effective 
pairs in the wave function with $N_{pp}={1\over 2}$ and $N_{pn}=4$.

Equations.(\ref{hrm_crs}) and (\ref{M2_fct}) together with Eqs.(\ref{phis}),(\ref{S12}),and (\ref{S34}) allow us to 
calculate the  differential cross section of both $pp$ and $pn$ breakup reactions off the   
$^3$He target. 
Notice  that, on the qualitative level, as it follows from Eqs.(\ref{hrm_crs}) and (\ref{M2_fct}) 
in the limit of  $s\gg M_{^3\textnormal{\scriptsize {He}}}^2$ and $s_{NN}\gg m_N^2$,  the HRM predicts an $s^{-11}$ invariant energy 
dependence of the differential cross section provided that the $NN$ cross section scales as $s^{-10}$.
However the numerical calculations of Eq.(\ref{M2_fct}) require a knowledge of 
the $NN$ helicity amplitudes at high energy and momentum transfers. Our strategy is  
to use Eq.(\ref{crs_NN}) to express $NN$ breakup reactions 
directly through the differential cross section of $pn$ and $pp$ elastic scatterings rather than to use 
helicity amplitudes explicitly.

\section{Hard breakup of proton and neutron from {\boldmath $^3$He}.}
\label{III}
We consider now the reaction
\begin{equation}
\gamma + ^3\textnormal{He} \rightarrow (pn) + p,
\label{pn_hrm}
\end{equation}
in which one proton is very energetic and produced at large c.m. angles with 
the neutron, while the second proton emerges with low momentum $\lesssim$ $100$~MeV/$c$. In this 
case the hard rescattering happens in the $pn$ channel. Using the $\phi_3\approx \phi_4$ relation for 
hard $pn$ scattering amplitude (see, e.g., Refs \cite{FGST,BCL,RS}) for breakup amplitude of Eq.(\ref{M2_fct}) one 
obtains
\begin{eqnarray}
\bar {|{\cal M}|}^2 = & &  {(Q^{pn}_{F}e)^2 2(2\pi)^6\over 2 s^\prime_{NN}}
{1\over 2}\left\{2|\phi_5|^2 S_0 + (|\phi_1|^2+|\phi_2|^2)S_{12} + (|\phi_3|^2 + |\phi_4|^2)S_{34}\right\},
\label{M2_fct_pn}
\end{eqnarray}
where $Q^{(pn)}_{F}= Q^{p}_F + Q^{n}_F$ can be calculated using Eqs.(\ref{QNNQIM}) and (\ref{QF}). Based on 
SU(6) flavor-spin symmetry of nucleon wave functions, for the helicity amplitudes  of Eq.(\ref{phis}) using 
the method described in Appendix C one obtains
\begin{equation}
Q^{pn}_{F} = {1\over 3}.
\label{QFpn}
\end{equation}
We can further simplify Eq.(\ref{M2_fct_pn}) noticing that 
for  the $pn$ pair in $^3$He  one has $S_{12}\approx S_{34} \approx {S_0\over 2}$. 
This is due to the fact that 
in the dominant $S$ state  two protons have opposite spins and therefore the probability of finding one 
proton with a helicity opposite to that of  the neutron is equal to the other proton 
having the same helicity as the neutron's. 
Using this relation and Eq.(\ref{crs_NN}) for the $pn$ breakup reaction one 
obtains
\begin{eqnarray}
\mid \bar M \mid^2 =  {(eQ_{F,pn})^2(2\pi)^6 \over s_{NN}^\prime}
16\pi s_{NN}(s_{NN}-4m_N^2) {d\sigma^{pn\rightarrow pn}(s_{NN},t_N)\over dt_N }{S^{pn}_0\over 2}.
\label{M2_pn}
\end{eqnarray}
Inserting it in Eq.(\ref{hrm_crs}) for the differential cross section one obtains
\begin{eqnarray}
{d\sigma^{\gamma ^3He\rightarrow (pn) p}\over dt {d^3p_s\over{E_s}}} =  
{\alpha Q^2_{F,pn} 16\pi^4}{S^{pn}_0(\alpha={1\over 2},\vec p_s)\over 2}
{s_{NN}(s_{NN}-4m^2)\over (s_{NN}-p^2_{NN})^2_{NN}(s-M^2_{^3\textnormal{\scriptsize {He}}})}{d\sigma^{pn\rightarrow pn}(s_{NN},t_N)\over dt_N },
\nonumber \\
\label{pn_hrm_crs}
\end{eqnarray}
where $\alpha = {1\over 137}$ and ${d\sigma^{pn \rightarrow pn}\over dt_N}$ is the differential cross 
section of hard $pn$ scattering 
evaluated at values of $s_{NN}$  and $t_N$ defined in Eq.(\ref{stuN}). The spectral function $S_0^{pn}$ 
is defined in Eq.(\ref{S0}) and corresponds to:
\begin{eqnarray}
S_0^{pn}(\alpha,\vec p_s) = 4\sum\limits_{\lambda_1,\lambda_2,\lambda_3=-{1\over 2}}^{1\over 2}
\left|\int \Psi_{^3\textnormal{\scriptsize {He,NR}}}^{1\over 2}
(\vec p_1,\lambda_1,{1\over 2};\vec p_2,\lambda_{2},-{1\over 2};\vec p_s,\lambda_3)
m_N{d^2p_{\perp} \over (2\pi)^2}\right|^2.
\label{S0pn}
\end{eqnarray}

\section{Hard breakup of two protons from {\boldmath $^3$He}}
\label{IV}
We now consider the reaction
\begin{equation}
\gamma + ^3\textnormal{He} \rightarrow (pp) + n,
\label{pp_hrm}
\end{equation}
in which two protons are produced at large c.m. angles while the neutron emerges 
as a spectator with small momentum ($p_s \le 100$~MeV/$c$).

We observe now that the relation between $S_{12}$ and $S_{34}$ is
very different from that of the  $pn$ case. Due to the fact that two protons 
cannot have the same helicity in the $S$ state one has $S_{12} \ll S_{34}$. 
The estimates of the spectral functions 
based on the realistic $^3$He wave function\cite{Nogga} gives ${S_{12}\over S_{34}}\sim 10^{-4}$. 
Therefore one can neglect the $S_{12}$ term in Eq.(\ref{M2_fct}).  
The next observation is that for  $pp$ scattering the helicity amplitudes 
$\phi_3$ and $\phi_4$ have opposite signs due to the Pauli principle (see, e.g., Refs.\cite{JW,FGST}).
Using the above observations and neglecting 
the  helicity-nonconserving amplitude $\phi_5$  for the $pp$-breakup amplitude we obtain
\begin{eqnarray}
\bar {|{\cal M} |}^2 = & &  {(e^2 2(2\pi)^6\over 2 s^\prime_{NN}}
{1\over 2}\left\{ 2(Q_F^{p}|\phi_3| - Q_F^{p}|\phi_4|)^2S_{34}\right\}.
\label{M2_fct_pp}
\end{eqnarray}
The charge factor $Q_F^p$ depends on the helicity amplitude it couples; therefore one estimates it 
for the combination of $(Q_F^{p}|\phi_3| - Q_F^{p}|\phi_4|)$. Using SU(6) symmetry for the distribution of 
given helicity-flavor valence quarks in the proton and through the approach described in Appendix C  we obtain
\begin{equation}
(Q_F^{p}|\phi_3| - Q_F^{p}|\phi_4|) = Q^{pp}_{F}(|\phi_3| - |\phi_4|)
\end{equation}
with 
\begin{equation}
Q_F^{pp} = {5\over 3}.
\label{QFpp}
\end{equation}
It is worth noticing that due to explicit consideration of quark degrees of freedom the 
effective charge involved in the breakup is larger for the case of two protons than for 
proton and neutron [see Eq.(\ref{QFpn})]. This is characteristic of the HRM model in which a 
photon couples to a  quark and more charges are exchanged in the $pp$ case than in the  $pn$
case. This is rather opposite to the scattering picture  considered based on 
hadronic degrees of freedom in which case the photon will couple to an exchanged meson and 
$pp$ contribution will be significantly suppressed because no charged mesons can be exchanged 
within the $pp$ pair.

To be able to estimate the cross section of the $pp$ breakup reaction through the 
elastic $pp$ scattering cross section we introduce a parameter 
\begin{equation}
C^2 = {\phi_3^2\over \phi_1^2}\approx {\phi_4^2\over \phi_1^2},
\label{cfactor}
\end{equation}
which allows to express the differential cross section of the reaction (\ref{pp_hrm}) in the following form:
\begin{eqnarray}
{d\sigma^{\gamma^3He\rightarrow(pp)n}\over dt {d^3p_{s}\over E_s}} = & &   
\alpha Q_{F,pp}^2 16\pi^4 S^{pp}_{34}(\alpha={1\over 2},\vec p_s){2\beta^2\over 1+2C^2}
{s_{NN}(s_{NN}-4m_N^2)\over (s_{NN}-p_{NN}^2)^2 (s-M_{^3\textnormal{\scriptsize {He}}}^2)}\times \nonumber \\
& & {d\sigma^{pp\rightarrow pp}(s_{NN},t_{N})\over dt},
\label{pp_hrm_crs}
\end{eqnarray}
where we also introduced a factor $\beta$,
\begin{equation}
\beta  = {|\phi_3| - |\phi_4|\over |\phi_1|},
\label{betafactor}
\end{equation}
which accounts for the suppression due to the cancellation between $\phi_3$ and $\phi_4$ 
helicity amplitudes of elastic $pp$ scattering.\footnote{This cancellation was overlooked in our early estimate 
of the cross section of $pp$ photodisintegration from $^3$He target 
(see, e.g., \cite{gheppn}).}
The spectral function $S^{pp}_{34}$ in Eq.(\ref{pp_hrm_crs}) is expressed through the $^3$He wave function
according to Eq.(\ref{S34})  as follows:
\begin{eqnarray}
S^{pp}_{34}(\alpha,\vec p_s) = {1\over 2} \sum\limits_{\lambda_1=-\lambda_2=-{1\over2}}^{1\over 2}
\sum\limits_{\lambda_3 
= -{1\over2}}^{1\over 2}\left| \int \Psi_{^3\textnormal{\scriptsize {He,NR}}}^{1\over 2}
(\vec p_1,\lambda_1,{1\over 2};\vec p_2,\lambda_{2},{1\over 2};\vec p_s,\lambda_3)
m_N{d^2p_{2,\perp} \over (2\pi)^2}\right|^2.
\label{S34pp}
\end{eqnarray}

\section{Two- and three-body processes in $NN$ breakup reactions}
\label{V}

For a two-body hard $NN$ breakup mechanism to be observed  it must 
dominate the three-body/two-step processes.  This is especially important for 
$pp$ breakup processes, (\ref{pp_hrm}) because according to Eqs.(\ref{pp_hrm_crs} and \ref{betafactor}) 
the  two-body contribution is suppressed because of a cancellation between 
$\phi_3$ and $\phi_4$ helicity amplitudes.

At low to intermediate range energies~($E_\gamma \sim 200$~MeV) it is rather well established that 
the $pp$ breakup reaction proceeds 
overwhelmingly through a two-step (three-body) process\cite{Laget,Laget1,Laget2,Jan1,Jan2,Giusti} 
in which the initial breakup 
of the $pn$ pair (dominated by $\pi^\pm$ exchange) is  followed by a charge-interchange final state 
interaction of the neutron with the spectator proton.  
Other two-step processes include the excitation of intermediate 
$\Delta$ isobars in the $pn$ system with the subsequent rescattering off the 
spectator neutron, which produces two final protons.

\begin{figure}[t]
\centering\includegraphics[height=5.cm,width=10.cm]{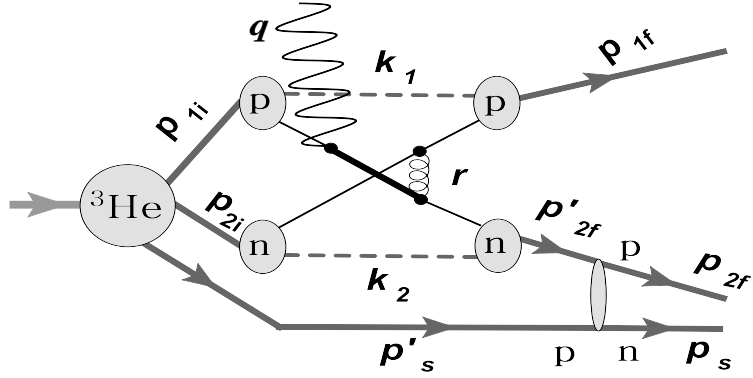}
\caption{(Color online) Diagram corresponding to 
three-body processes in which the hard breakup of the $pn$ pair is followed by a soft charge-exchange  
rescattering of the neutron off the spectator proton.}
\label{3bfig}
\end{figure}

The dominance of three-body processes at low energies is related mainly to the fact that 
the two-body $pp$ breakup is negligible because of the impossibility of charged-pion 
exchanges between two protons that absorb an incoming photon.

At high energy kinematics within the HRM the interaction between protons is carried out by exchanged quarks
because of which the relative strength of $pp$ breakup is larger. 

To estimate the strength of three-body processes  at high energy kinematics, one needs to calculate 
the contribution of diagrams similar to Fig. \ref{3bfig}.  Because the charge-exchange rescattering 
at the final stage of the process in Fig.\ref{3bfig} takes place at proton momenta 
$p_{f}^\prime > m_N$, one 
can apply an eikonal approximation\cite{gea,stopics} to estimate its contribution.

For $E_\gamma \ge 2$~GeV assuming that the HRM is valid for the first ($pn$ breakup) stage of the reaction, 
for the amplitude of three-body/two-step process within the eikonal approximation\cite{gea,stopics} one obtains
\begin{eqnarray}
M_{3body} & \approx &  
{eQ_{F,pn}(2\pi)^3\over 2\sqrt{2s_{NN}^\prime}} T^{hard}_{pn\rightarrow pn}(t_N) 
\times \nonumber \\
& &  \int  \Psi_{^3\textnormal{\scriptsize {He,NR}}}^{\lambda_A}
(\vec p_1,t_1;\vec p_2,t_2;\vec p_s-\vec k_\perp)
m_N {T^{chex}_{pn\rightarrow np}(k_\perp)\over s_{NN}}{d^2p_{\perp} \over (2\pi)^2}{d^2k_{\perp} \over (2\pi)^2},
\label{3b}
\end{eqnarray}
where we suppressed helicity indices for simplicity and choose the isospin projections,  
$t_{1} = -t_{2} = {1\over 2}$, corresponding to the initial $pn$ pair that interacts with the photon. Here 
$T^{\textnormal{\scriptsize hard}}_{pn\rightarrow pn}(t_N)$ is the hard elastic $pn$ scattering amplitude and 
$T^{\textnormal{\scriptsize chex}}_{pn\rightarrow np}$ represents  the amplitude of the soft charge-exchange $pn$ scattering.
Because of the pion-exchange nature of the latter it is rather well established that this amplitude 
is real and can be represented as $\sim$$\sqrt{s}A e^{{B\over 2}t}$, where $A$ and $B$ are approximately
constants\cite{GL}.

Two main  observations follow from Eq.(\ref{3b}) and the above-mentioned property of the charge-exchange amplitude:
First, three-body and two-body amplitudes [see, e.g., Eq.(\ref{ampl5})] will  not interfere, since one is real and 
the other is imaginary.  The fact that these two amplitudes differ by order of $i$ follows from the general 
structure of rescattering amplitudes (see, e.g., Ref.\cite{stopics}).  Equation (\ref{ampl5}) corresponds to  a  
single rescattering amplitude,  while Eq.(\ref{3b}) to a  double rescattering amplitude.\\
Second, because of the energy dependence of the charge-exchange scattering amplitude at small angles,  
the three-body contribution will scale like $s^{-12}$ as compared to the 
two-body breakup contribution.

Using Eq.(\ref{3b}) one can estimate the magnitude of the contribution of three-body processes in the 
$pp$ breakup cross section as follows:
\begin{equation}
{d\sigma^{\gamma^3He\rightarrow(pp)n}_{three-body}\over dt {d^3p_{s}\over E_s}} \approx 
{d\sigma^{\gamma ^3He\rightarrow (pn) p}_{two-body}\over dt {d^3p_s\over{E_s}}}{S^{pnp}(p_s)
\over S_{0}^{pn}(p_s)}
\label{3to2}
\end{equation}
where $S_{0}^{pn}(p_s)$ is defined in Eq.(\ref{S0pn}) and for 
$S^{pnp}(p_s)$ based on Eq.(\ref{3b}) one obtains
\begin{equation}
S^{pnp}(p_s) = {N_{pn}\over 16 s_{NN}^2} \mid \int  \Psi_{^3\textnormal{\scriptsize {He,NR}}}^{\lambda_A}
(\vec p_1,\vec p_2,\vec p_s-\vec k_\perp)m_N T^{chex}_{pn\rightarrow np}(k_\perp)
{d^2p_{\perp} \over (2\pi)^2}{d^2k_{\perp} \over (2\pi)^2}\mid^2.
\label{spnp}
\end{equation}
Here both spectral functions are defined at $\alpha={1\over 2}$.

Using Eqs.(\ref{3to2}) and (\ref{spnp}) and the parametrization of $T^{\textnormal{\scriptsize chex}}_{pn\rightarrow np}$ from 
Ref.\cite{GL} one can estimate the relative contribution of three-body processes numerically. Note that this 
contribution is maximal at $\alpha_s=1$ and  increases with an increase of the momentum of $p_s$. However 
because of the charge-exchange  nature of the second  rescattering, this contribution decreases linearly 
with an increase of $s$.\footnote{Notice that for the case of diagonal $pn\rightarrow pn$  rescattering  
$T_{pn\rightarrow np}(k_\perp)= s\sigma_{\textnormal{\scriptsize tot}}e^{{B\over 2}t}$ and as a result the 
probability of the rescattering is energy independent.}

\section{Numerical Estimates}
\label{VI}
For numerical estimates we consider the center of mass reference frame of $\gamma$$NN$ system, for which 
according to Eq.(\ref{mans_st}) one obtains
\begin{equation}
t_{N,N} = - {(s_{NN}-M_{NN}^2)\over 2\sqrt{s_{NN}}}(\sqrt{s_{NN}}
-\sqrt{s_{NN}-4 m_N^2}cos(\theta_{\textnormal{\scriptsize c.m.}})) + m_N^2,
\label{tgnn_cm}
\end{equation}
where $M_{NN}^2 = p^\mu_{NN} p_{NN,\mu}$ and $p^\mu_{NN} = p^\mu_{^3\textnormal{\scriptsize {He}}}-p^\mu_s$. Using the above equation
we obtain for $t_{N}$ [Eq.(\ref{stuN})], which defines the effective momentum transfer in the $NN$ scattering
amplitude, the following relation: 
\begin{equation}
t_{N} =  - {(s_{NN}-M_{NN}^2)\over 4\sqrt{s_{NN}}}(\sqrt{s_{NN}}
-\sqrt{s_{NN}-4 m_N^2}cos(\theta_{\textnormal{\scriptsize c.m.}})) + m_N^2 - {M_{NN}^2\over 4}.
\label{tN}
\end{equation}
One can also calculate the effective c.m. angle that enters in the $NN$ scattering amplitude as follows:
\begin{equation}
cos(\theta_{\textnormal{\scriptsize c.m.}}^N) = 1 - {(s_{NN}-M_{NN}^2)\over 2(s_{NN}-4m_N^2)}
{(\sqrt{s_{NN}}-\sqrt{s_{NN}-4 m_N^2}cos(\theta_{\textnormal{\scriptsize c.m.}}))\over \sqrt{s_{NN}}} + 
{4m_N^2 - M_{NN}^2\over 2(s_{NN}-4m_N^2)}.
\label{theta_cmn}
\end{equation}
The above equations define the kinematics of hard $NN$ rescattering. 
\begin{figure}[t]
\centering\includegraphics[height=8.6cm,width=8.6cm]{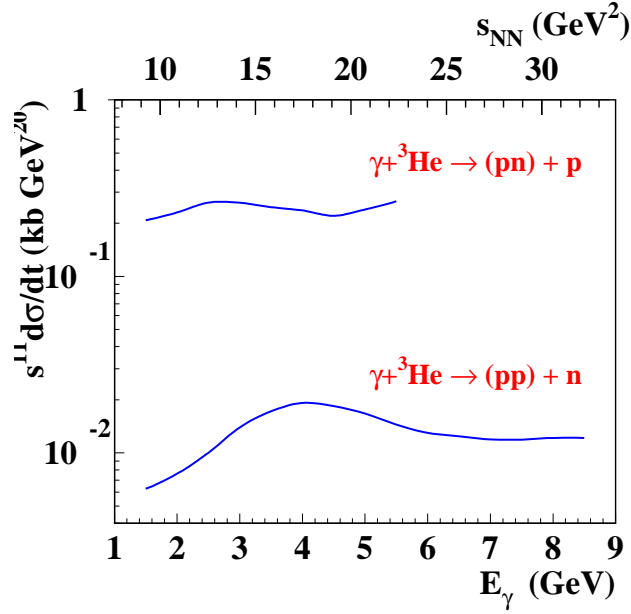}
\caption{(Color online). Energy dependence of $s^{11}$ weighted differential cross sections at 
$90^\textnormal{\scriptsize o}$ c.m. angle scattering in "$\gamma$-$NN$"system. In these calculations one integrated over 
the spectator nucleon momenta in the range of $0$-$100$~MeV/$c$.}
\label{Fig.s_dep}
\end{figure}

\subsubsection{Energy dependence and the magnitude of the cross sections}

We are interested in energy dependences of the hard breakup reactions of Eqs.(\ref{pn_hrm}) and 
(\ref{pp_hrm}) at fixed and large angle production of two fast nucleons in the ``$\gamma$-$NN$'' center of 
mass reference frame. Particularly interesting is the case of $\theta_{\textnormal{\scriptsize c.m.}}$=$90^{\textnormal{\scriptsize 0}}$ for which 
as it follows from Eq.(\ref{theta_cmn}) $cos(\theta_{\textnormal{\scriptsize c.m.}}^N) = 0.5$.   This means that the 
cross sections of hard breakup reactions at these kinematics will be defined by the $NN$ elastic 
scattering at $\theta^N_{\textnormal{\scriptsize c.m.}} = 60^{\textnormal{\scriptsize 0}}$. In Fig.\ref{Fig.s_dep} the $E_\gamma$ and $s$ dependencies of the  
$s_{NN}^{11}$ weighted differential cross sections are presented for the cases of the $pp$ and $pn$ 
breakup reactions. In the calculation we integrated over the spectator nucleon's momentum  
in the range of (0-100)~MeV/$c$ and over the whole range of its solid angle.
Also for the parameter $C$ in Eq.(\ref{cfactor}) we used $C={1\over 2}$, consistent with an estimate 
obtained within the quark-interchange model of $pp$ scattering (see, e.g.,\cite{FGST,BCL}). The estimation of 
the factor $\beta$, which takes into account the cancellation between $\phi_3$ and $\phi_4$ helicity 
amplitudes in Eq.(\ref{pp_hrm_crs}) requires the knowledge of the angular dependence for 
helicity amplitudes. For this we used the helicity amplitudes calculated within 
thequark-interchange model\cite{BCL,FGST} with phenomenological angular dependencies estimated using  
$F(\theta_{\textnormal{\scriptsize c.m.}}) = {1\over sin^2(\theta_{cm})(1-cos(\theta_{cm}))^2}$ 
(see, e.g., Refs. \cite{BCL,RS}) which  describes reasonably well the data at hard scattering kinematics.

\begin{figure}[t]
\centering\includegraphics[height=8.6cm,width=8.6cm]{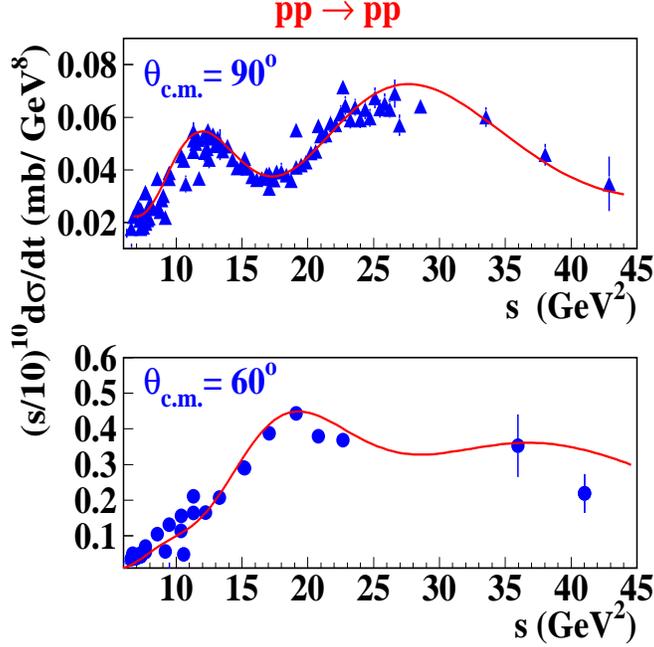}
\caption{(Color online) Invariant energy dependence of $s^{10}$ weighted differential cross sections of 
elastic $pp$ scattering at $\theta_{\textnormal{\scriptsize c.m.}}=90^{\textnormal{\scriptsize o}}$ and $\theta_{\textnormal{\scriptsize c.m.}}=60^{\textnormal{\scriptsize o}}$. Curves 
are the  fits~\cite{Carlos} of the available world data~\cite{hepdata}.}
\label{Fig.pp}
\end{figure}

\begin{figure}[ht]
\centering\includegraphics[height=8.6cm,width=8.6cm]{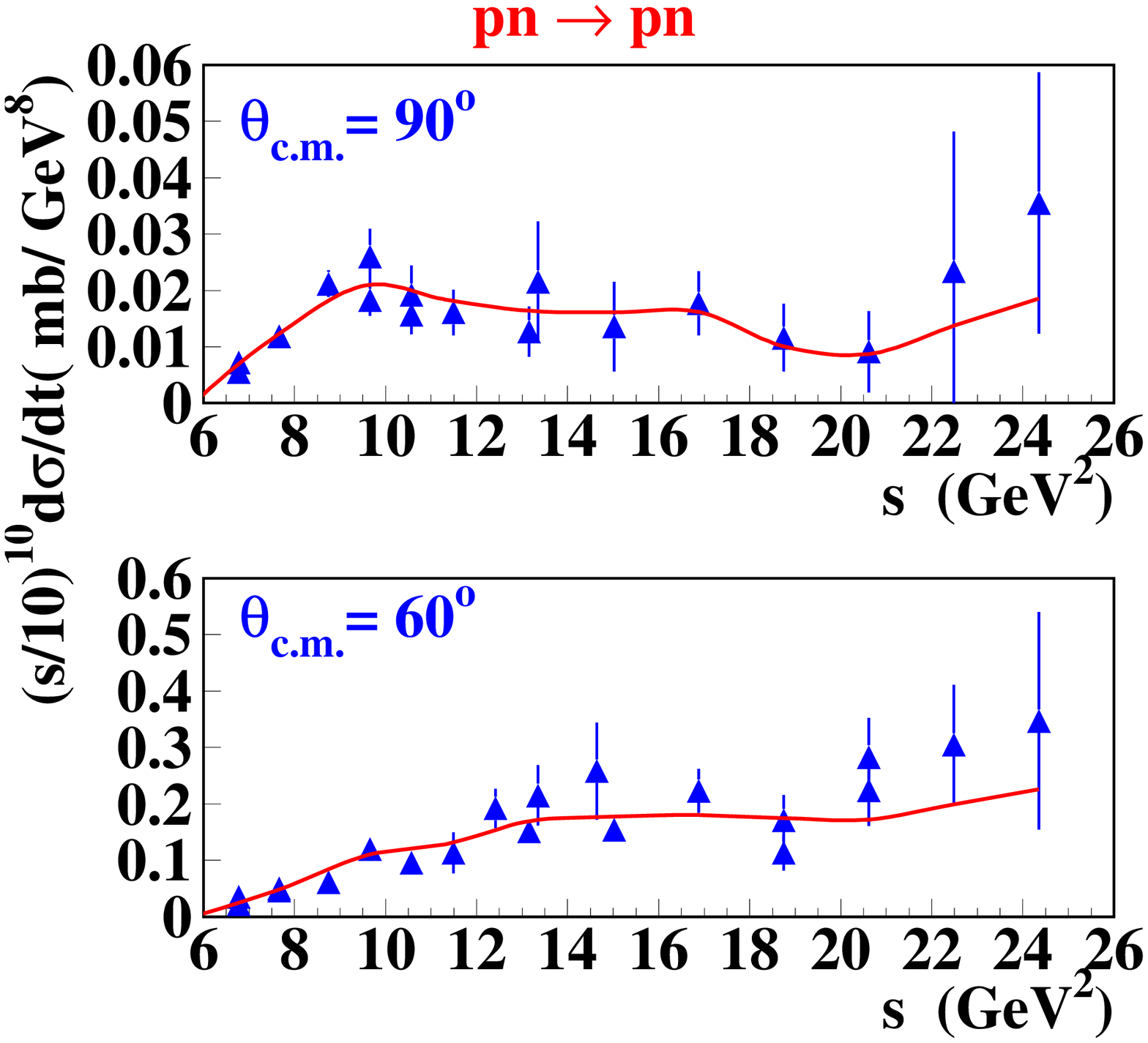}
\caption{(Color online) Invariant energy dependence of $s^{10}$ weighted differential cross sections of 
elastic $pn$ scattering at $\theta_{\textnormal{\scriptsize c.m.}}=90^{\textnormal{\scriptsize o}}$ and $\theta_{\textnormal{\scriptsize c.m.}}=60^{\textnormal{\scriptsize o}}$. Curves 
are the  fits~\cite{Carlos} of the available world data~\cite{hepdata}.}

\label{Fig.pn}
\end{figure}

Several features of the HRM  calculations are worth discussing in Fig.\ref{Fig.s_dep}: First, the breakup cross sections in 
average scale like $s_{NN}^{-11}$. Note that the absolute~(nonscaled) values of the cross sections  drop by 
five orders of magnitude in the 2-8~GeV of photon energy range.
Next, the shapes of the $s^{11}$ weighted differential 
cross sections reflect the shapes of the $s^{10}$ weighted differential cross sections of $pp$ and $pn$
scattering at $\theta_{\textnormal{\scriptsize c.m.}}=60^{\textnormal{\scriptsize o}}$. [see Figs.(\ref{Fig.pp}) and (\ref{Fig.pn})]. It 
is worth noting that as follows from  Figs.(\ref{Fig.pp}) and (\ref{Fig.pn}) 
the fits used in the calculation of $pp$ and $pn$ breakup reactions contain uncertainties
on the level of 10\% for $pp$ breakup (for $s_{NN}\ge 24$~GeV$^2$) and up to 30\% 
for $pn$ breakup reactions.  Based on this, one can conclude that the calculated shape of the energy 
dependence of the $pn$ breakup reaction in Fig.(\ref{Fig.s_dep}) does not have much predictive power.
However, for the  $pp$ breakup the calculated shape, for up to $s_{NN}\le 24$~GeV$^2$, is not 
obscured by the uncertainty of the $pp$ data and can be considered as a prediction of the HRM.
It is worth mentioning that considered features of the HRM  
are insensitive to the choice of the above-discussed parameters of $C$ and $\beta$, because they only 
define the absolute magnitude of the $pp$ breakup cross section.

The next feature of the calculations in Fig.\ref{Fig.s_dep} is the magnitude of 
the $pn$ and $pp$  breakup cross sections. The $pn$ breakup cross section 
[Eq.(\ref{pn_hrm_crs})] does not contain any free parameter, and similar
to the HRM prediction for the breakup  of the deuteron\cite{gdpn}, 
it is expressed through the rather well-defined 
quantities.  For the estimate of $pp$ breakup, however, one needs to know the  relative strength of 
the $\phi_3$ and $\phi_4$ amplitudes as compared to $\phi_1$ as well as the extent of their 
cancellation at kinematics of $s_{NN}$ and $t_N$ defined in Eqs.(\ref{stuN}) and (\ref{tN}).	
Our calculation,  based on  phenomenologically justified estimates of factors $C$ and $\beta$ in 
Eq.(\ref{pp_hrm_crs}) results in the $pp$ breakup cross section which is  about ten times 
smaller than the cross section for the $pn$ breakup.  This magnitude indicates however, 
an  increase of $pp$ breakup cross section relative to the $pn$ breakup cross section as 
compared to the results from the low energy breakup reactions.
As it was mentioned in Sec.\ref{V}  
at low energies ($\sim$ 200~MeV) the cross section of $pp$ photodisintegration from $^3$He is 
significantly  smaller (by almost two orders of magnitude according to Ref.\cite{Laget}) than 
the $pn$-breakup cross section.  

Note that the  factors $C$ and $\beta$ introduce an additional uncertainty in the estimation of the magnitude 
of the $pp$-breakup cross section. While the factor $C$ can be evaluated  in  the quark-interchange model 
thus staying within the framework of the considered model, the factor $\beta$ is not constrained by 
the theoretical framework of the model.  The latter is sensitive to the angular dependence, $F(\theta_{c.m.})$, 
of the helicity amplitudes. To estimate the uncertainty due to $F(\theta_{c.m.})$ we varied it 
around the form,  $F(\theta_{\textnormal{\scriptsize c.m.}}) = {1\over sin^2(\theta_{\textnormal{\scriptsize c.m.}})(1-cos(\theta_{\textnormal{\scriptsize c.m.}}))^2}$ in such 
a way that the results were still in agreement with angular distribution of $pp$ scattering 
at $-t,-u > 1$~GeV$^2$. We found that this variation 
changes the HRM prediction for the magnitude of  $pp$-breakup cross section as much as $40\%$.

\begin{figure}[ht]
\centering\includegraphics[height=8.6cm,width=8.6cm]{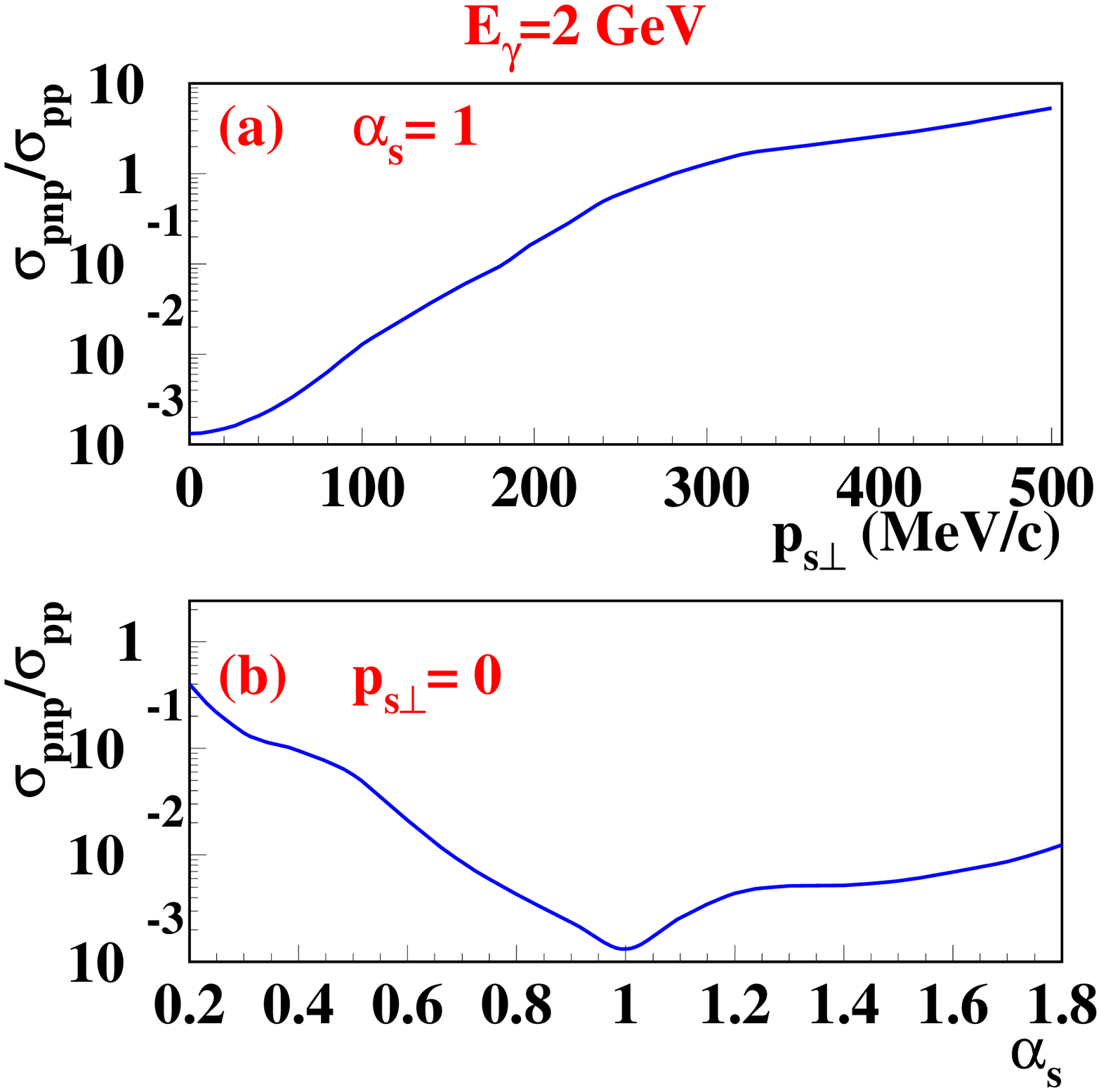}
\caption{Dependence of the ratio of the cross section of  
three-body/two-step process discussed in Sec.V to the cross section of the two-body $pp$ breakup  
at $E\gamma=2$~GeV on (a) transverse momentum of the spectator neutron $p_{st}$ at $\alpha_s=1$ and 
on (b) $\alpha_s$ at $p_{st}=0$.}
\label{Fig.3b_2b}
\end{figure}

Because $pp$ breakup cross section is still by a factor of 10 smaller than the $pn$ cross section,
one needs to estimate the contribution due to three-body processes in which hard $pn$ breakup 
is followed by soft charge-exchange rescattering.  The estimate based on Eq.(\ref{3to2}) is 
given in Fig. \ref{Fig.3b_2b} where the ratio of three-body  to two-body breakup cross 
sections is evaluated for different values of $\alpha_s$ and  transverse momentum  
of the spectator neutron, $p_{s\perp}$.

Because of the eikonal nature of the second rescattering in three-body processes, one expects the cross section 
to be maximal at $\alpha_s=1$. As Fig.\ref{Fig.3b_2b}(a) shows  in this case, the three-body process is 
a correction to the two-body breakup process, $\sim$2\% for $p_{s\perp}=100$~MeV/$c$ and $\sim 17$\% for 
$p_{s\perp}=200$~MeV/$c$. Then, starting at $p_{s\perp}$ > 300~MeV/$c$ the three-body  process dominates 
the two-body contribution.
The latter can be verified by observing an onset of $s^{-12}$ scaling at large ($\ge 300$~MeV/$c$) 
transverse momenta of the spectator neutron in the case of hard $pp$ breakup reactions.  
Fig.\ref{Fig.3b_2b}(b) shows also  that the three-body contribution will be always small for 
$p_{s\perp}\approx 0$~MeV/$c$, and for a wide range of $\alpha_s$,  which again reflects the eikonal nature of 
the second order rescattering in  which case the recoiling of the spectator  nucleons happens predominantly 
at $\sim 90^{\textnormal{\scriptsize o}}$ (see, e.g., \cite{Glauber}). Note that one expects the above estimate of the three-body 
contribution to contain an uncertainty of  10-15\%, representing the general level of accuracy of 
eikonal approximations.

Based on Fig.\ref{Fig.3b_2b} one can expect that overall, for small values of $p_{s}\le$100-150~MeV/$c$    
in the high energy limit ($E_\gamma > 2$~GeV) one expects two-body breakup mechanisms to dominate  for 
both $pp$ and $pn$ production reactions.

\subsection{Spectator nucleon momentum dependence}

The presence of a spectator nucleon in the hard two-nucleon breakup reaction from $^3$He 
gives us an additional degree of freedom in checking the mechanism of the photodisintegration.
As follows from Eqs.(\ref{pn_hrm_crs}) and (\ref{S0pn}) and Eqs.(\ref{pp_hrm_crs}) and (\ref{S34pp}) the $pn$ and $pp$ breakup
cross sections  within the HRM are sensitive to different components of the nuclear spectral function. This is due to 
the fact that the $pp$ component with the same helicities for both protons is suppressed in the ground state 
wave function of the $^3$He target.  Thus one expects rather different spectator-momentum dependencies 
for $pp$ and $pn$ breakup cross sections.

\begin{figure}[ht]
\centering\includegraphics[height=8.6cm,width=8.6cm]{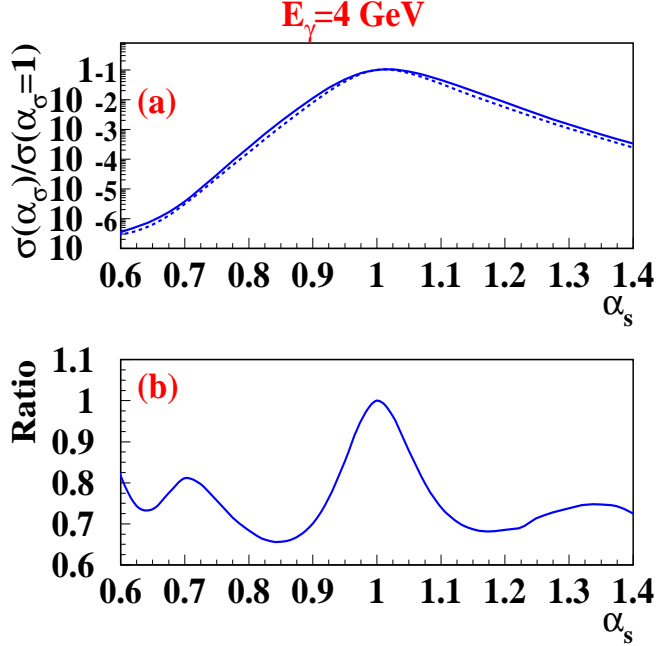}
\caption{(Color online). Dependence of the $s^{11}$ weighted 90$^{\textnormal{\scriptsize o}}$ c.m. breakup differential cross section
on the light-cone momentum fraction of  of 
spectator nucleon, $\alpha_s$,  calculated at $E_\gamma=4$~GeV and $p_{s\perp}=0$. 
(a) The solid line is for $pp$ breakup reactions, and the dashed line is for $pn$ breakup 
reactions. Calculations are normalized to the cross sections at $\alpha_s=1$. (b) Ratio of the $pn$ to $pp$ 
breakup cross sections normalized to their values at $\alpha_s=1$.}
\label{Fig.al_dep}
\end{figure}

The quantity that we consider for numerical estimates is not the momentum of the spectator but 
rather the  momentum fraction of the target carried by the spectator nucleon, $\alpha_s$.   This quantity is 
Lorentz invariant with respect to boosts in the $q$ direction, which allows us to specify it 
in the Lab frame as follows:
\begin{equation}
\alpha_{s} \equiv {E_s-p_{s,z}\over M_A/A} = \alpha_A - \alpha_{1f} - \alpha_{2f}
\label{alphas}
\end{equation}
where $\alpha_{i} = {E_i-p_{i,z}\over M_A/A}$ for $i=A$, $1f$, $2f$ and $z$ axis in the Lab frame is 
defined parallel to the momentum of incoming photon $q$.  Note that the photon does not 
contribute to the above equation because $\alpha_{q}=0$. In definition of $\alpha_s$ we use a normalization 
such that for a stationary spectator $\alpha_s=1$.
The $\alpha_s$ dependencies of the differential cross sections for $pp$ and $pn$ breakup reactions 
normalized to their values at $\alpha_s=1$ are given in Fig.\ref{Fig.al_dep}(a). One feature of $\alpha_s$
dependence is the asymmetry of the cross section around $\alpha_s=1$ with cross sections 
dominating at $\alpha_s>1$. This property can be understood from the fact that the momentum fraction of 
the $NN$ pair that breaks up is defined through $\alpha_s$ as follows:
\begin{equation}
\alpha_{NN} = 3-\alpha_{s}.
\end{equation} 
The latter quantity defines the invariant energy of the $NN$ pair as follows
\begin{equation}
s_{NN} = M_{NN}^2 + E_\gamma m_n\alpha_{NN}.
\end{equation}
Because the cross section within the HRM is proportional to $s_{NN}^{-10}$,\footnote{An additional negative power 
of invariant energy is provided by ${1\over s-M_{A}^2}$ factor  in the differential cross section of 
the reaction [see Eqs.(\ref{pn_hrm_crs},\ref{pp_hrm_crs})].} it will be enhanced at small values of $s_{NN}$ 
that will correspond to smaller values of $\alpha_{NN}$ or larger values of $\alpha_s$.

The difference of the cross sections due to the different composition of the nuclear spectral functions 
entering the $pp$ and $pn$ breakup reactions can be seen in Fig.\ref{Fig.al_dep}(b) in which 
case one 
calculates the ratio of $pn$ to $pp$ breakup cross sections normalized to their 
values at $\alpha_s=1$.   The drop of the ratio in Fig.\ref{Fig.al_dep}.(b) at values close to 
$\alpha_s=1$ is due to the suppression of the same-helicity two-proton 
component in the ground state nuclear wave function at small momenta. In this  case the  
spectral function is sensitive to the higher angular momentum components of the ground state 
nuclear wave function. This yields a  wider momentum distribution for the $pp$ spectral function 
as compared to that for $pn$ because no same-helicity state suppression exits for the latter.
The estimates indicate that differences in $\alpha_s$ dependencies of $pp$ and $pn$ breakup 
cross sections are rather large  and can play an additional role in checking the validity of 
the HRM.

\section{Polarization transfer of the hard rescattering  mechanism}
\label{VII}
One of the unique properties of the hard rescattering mechanism of two-nucleon breakup is 
that the  helicity of the nucleon from which a quark is 
struck is predominantly defined by the helicity of the incoming photon $\lambda_{1i} = \lambda_\gamma$ 
[see Eq.(\ref{ampl5})].  This is based on the fact that in the massless quark limit the helicity of 
the struck quark equals the helicity of the photon, $\eta_{1i}=\lambda_{\gamma}$, and assuming that at 
large $x$ the quark carries almost all the helicity of the parent nucleon one obtains 
$\lambda_{1i}\approx \eta_{1i}=\lambda_{\gamma}$.

Because within the HRM, the   energetic struck quark shares its momentum with a quark of the 
other nucleon through a hard gluon exchange, it will retain its initial helicity when it merges into 
the final outgoing nucleon.   It will also have $x^\prime\sim x\sim 1$, which allows 
us to conclude that the final outgoing nucleon will acquire the large part of struck quark's (as well as the 
photon's) helicity.  This mechanism will result in a large (photon) polarization transfer  for the 
hard two-nucleon breakup reactions.

An observable that is sensitive to polarization transfer processes 
is the quantity $C_{z^\prime}$, which for a circularly polarized photon measures the 
asymmetry of the hard breakup reaction with respect to the helicity of the 
outgoing proton.

A large value of $C_{z^\prime}$ was predicted within the HRM for the hard breakup of the deuteron 
in Ref.\cite{gdpnpol} that was observed in the  recent experiment of Ref.\cite{gdpnpolexp2}.   

For the case of the $^3$He target an additional experimental observation will be a comparison 
of $C_{z^\prime}$ asymmetries for $pp$ and $pn$ breakup channels. 
For the $^3$He target we define $C_{z^\prime}$ as follows:
\begin{eqnarray}
C_{z^\prime}  = 
{\sum\limits_{\lambda_{2f},\lambda_s,\lambda_a}
\left\{ 
\left|\langle +,\lambda_{2f},\lambda_s\mid M \mid +, \lambda_A\rangle\right|^2 - 
\left|\langle -,\lambda_{2f},\lambda_s\mid M \mid +, \lambda_A\rangle\right|^2\right\}\over
\sum\limits_{\lambda_{1f}\lambda_{2f},\lambda_s,\lambda_a}
\left|\langle \lambda_{1f},\lambda_{2f},\lambda_s\mid M \mid +, \lambda_A\rangle\right|^2}.
\label{Cz1}
\end{eqnarray}
Using Eq.(\ref{ampl5}) and the definitions of Eq.(\ref{phis}) for $C_{z^\prime}$ one obtains
\begin{equation}
C_{z^\prime} = {(|\phi_1|^2 - |\phi_2|^2)S^{++}+ (|\phi_3|^2 - |\phi_4|^2)S^{+-}\over
2|\phi_5|^2S^+ + (|\phi_1|^2 + |\phi_2|^2)S^{++}+ (|\phi_3|^2 + |\phi_4|^2)S^{+-}},
\label{Cz2}
\end{equation}
where
\begin{eqnarray}
& & S^{\pm,\pm}(t_1,t_2,\alpha,\vec p_s) = \nonumber \\ 
& & \ \ \ \ \sum\limits_{\lambda_A=-{1\over 2}}^{1\over 2}
\sum\limits_{\lambda_3 
= -{1\over2}}^{1\over 2} \left|\int \Psi_{^3\textnormal{\scriptsize {He,NR}}}^{\lambda_A}
(\vec p_1,\lambda_1=\pm{1\over 2},t_1;\vec p_2,\lambda_{2}=\pm{1\over 2},t_2;\vec p_s,\lambda_3)
m_N{d^2p_{2,\perp} \over (2\pi)^2}\right|^2
\label{Spm}
\end{eqnarray}
and $S^+ = S^{++} + S^{+-}$.

\medskip
\medskip

As follows from Eqs.(\ref{Cz2}) and (\ref{Spm}) one predicts significantly different 
magnitudes for $C_{z^\prime}$ for $pp$ and $pn$ breakup cases.

For the $pp$ breakup, $S_{pp}^{++}\ll S_{pp}^{+-}$ due to the smallness of the nuclear wave function component 
containing two protons in the same helicity state. As a result one expects
\begin{equation}
C_{z^\prime}^{pp} \approx {|\phi_3|^2-|\phi_4|^2\over |\phi_3|^2+|\phi_4|^2} \sim 0,
\label{czpp}
\end{equation}
while for the $pn$ break up case $S_{pn}^{++}\approx S_{pn}^{+-}$ then one obtains
\begin{equation}
C_{z^\prime}^{pn} \approx {|\phi_1|^2 + |\phi_3|^2-|\phi_4|^2\over 
|\phi_1|^2 + |\phi_3|^2+|\phi_4|^2} \sim {2\over 3},
\label{czpn}
\end{equation}
where in the last part of the equation we assumed that $|\phi_3| = |\phi_4| = {1\over 2}|\phi_{1}|$.

\section{Conclusions}
\label{VIII}
The hard rescattering mechanism of a two-nucleon breakup from the $^3$He nucleus at large c.m. angles
is based on the assumption of  the dominance of quark-gluon degrees of freedom in the hard scattering 
process involving two nucleons.
The model explicitly assumes that the  photodisintegration process proceeds through the 
knock-out of a quark from one nucleon with a subsequent rescattering of that quark 
with a quark from the second nucleon.  While photon-quark scattering is calculated explicitly, 
the sum of all possible quark rescatterings is related to the hard elastic 
$NN$ scattering amplitude. Such a relation is found assuming that quark-interchange 
amplitudes give the dominant contribution in the  hard elastic $NN$ scattering.

The model allows one to calculate the cross sections of the hard breakup of $pn$ and $pp$ pairs from $^3$He
expressing them through the amplitudes of elastic $pn$ and $pp$ scatterings, respectively.

Several results of the HRM are worth mentioning: First, the HRM predicts an approximate $s^{-11}$ scaling 
consistent with the predictions of the quark-counting rule. However, the model by itself 
is nonperturbative because the bulk of the incalculable part of the scattering amplitude is hidden 
in the amplitude of the $NN$ scattering that is taken from the experiment.

Second, because the hard $NN$ scattering amplitude enters into the  final amplitude of the photodisintegration 
reaction, the shape of the energy dependence of the $s^{11}$ weighted breakup cross section reflects the shape of 
the $s^{10}$ weighted $NN$ elastic scattering cross section. Because of a better accuracy of $pp$ elastic scattering 
data for $s_{NN}\le 24$~GeV$^2$,  we are able to predict a specific shape for the energy dependence of 
the hard $pp$ breakup cross section at photon energies up to $E_\gamma \le 5$~GeV.

Another observations is that, when $s^{-11}$ scaling is established, the HRM predicts an increase of the strength 
of the $pp$ breakup cross section relative to the $pn$ breakup as compared to the 
low energy results. This is due to the feature  that 
within the quark-interchange mechanism of $NN$ scattering one has more charges flowing between nucleons 
in the $pp$ pair than in the $pn$ pair. This situation is  opposite in the low energy regime 
when no charged meson exchanges exist for the $pp$ pair.  Even though the large charge factor is involved 
in the $pp$ breakup its cross section is still by a factor of ten smaller than the cross section of 
the $pn$ breakup. Within the HRM, this is due to cancellation between the helicity conserving 
amplitudes $\phi_3$ and $\phi_4$, which have opposite signs for the $pp$ scattering.

Because of the smallness of the $pp$ breakup cross section, within the eikonal approximation, 
we estimated the possible contribution of 
three-body/two-step processes in which the initial two-body hard $pn$ breakup is followed by the 
charge-exchange rescattering of an energetic neutron off the spectator proton.  We found that this 
contribution has $s^{-12}$ energy dependence and is a small correction for spectator nucleon 
momenta $\le$150~MeV/$c$. However, the three-body/two-step process will dominate the hard $pp$ breakup 
contribution at large  transverse momenta of the spectator nucleon starting at $p_{s\perp}\ge 350$~MeV/$c$.

The next result of the HRM is the prediction of different spectator-momentum dependencies of breakup 
cross section for the $pp$ and $pn$ pairs.  This result follows from the fact that the ground state 
wave function of $^3$He containing two protons with the same helicity is significantly suppressed as 
compared to the same component in the $pn$ pair. Because of this,  the $pp$ spectral function 
is sensitive to the higher angular momentum components of the nuclear ground state wave function. 
These components generate wider momentum distribution as compared to say the $S$ component of the 
wave function.  As a result the cross section of the $pp$ breakup reaction exhibits wider momentum 
distribution as compared to the $pn$ cross section. 
Additionally because of the strong $s$ dependence of the reaction, the cross section exhibits an 
asymmetry in the light-cone momentum distribution of the spectator nucleon, favoring larger values of 
$\alpha_s$.

The final result of the HRM is the strong difference in prediction of the polarization transfer asymmetry 
for $pp$ and $pn$ breakup reactions for circularly polarized photons.
Because of the suppression of the same helicity $pp$ components in the $^3$He ground state wave function, the 
dominant helicity conserving $\phi_1$ component will not contribute to the polarization transfer 
process involving two protons.  Because of this effect,  the HRM predicts longitudinal polarization transfer  
$C_{z\prime}$, 
for the $pp$ breakup to be close to zero.  Because no such suppression exists for the $pn$ breakup, the HRM 
predicts a rather large magnitude for $C_{z'}\approx {2\over 3}$. 

Even though the HRM model does not contain free parameters,  for numerical estimates we use the  magnitude of 
elastic $NN$ cross sections as well as  some properties of the $NN$ helicity amplitudes.  This introduces certain 
error in our prediction of the magnitudes of the breakup cross sections. For the $pn$ breakup this error is mainly 
related to the uncertainty in the magnitude of the absolute cross section of hard elastic $pn$ scattering which 
is on the level 30\%.
For the $pp$ breakup the main source of the uncertainty is the magnitude of the  cancellation 
between $\phi_3$ and $\phi_4$, which is sensitive to the angular distribution of helicity amplitudes. 
The uncertainty due to the angular function is on the level of 40~\%.  These uncertainties should be 
considered on top of the theoretical uncertainties that the HRM contains due to approximations such 
as estimating the scattering amplitude at maximal value of the nuclear wave function at $\alpha = {1\over 2}$.
The latter may introduce an uncertainty as much as $20\%$ in the breakup cross section.

In conclusion we expect that experimental verification of all above-mentioned predictions 
may help us to  verify the validity of the hard rescattering model. Also,  the  progress in extracting the helicity 
amplitudes of the hard $NN$ scattering will allow us to improve the accuracy of the HRM predictions.

\acknowledgments
We are grateful to  Drs. Ronald Gilman, Leonid Frankfurt, Eli Piasetzky and Mark Strikman for 
many useful discussions and comments. Special thanks to Drs. Ronald Gilman, Eli Piasetzky 
and Ishay Pomerantz for the discussion of their experimental results  that allowed us to find 
the cancellation factor in the $pp$-breakup cross section, which was overlooked in our calculations earlier.
This work is supported by U.S. Department of Energy Grant 
under Contract DE-FG02-01ER41172.

\appendix

\section{Calculation of the {\boldmath $^3$He$(\gamma, NN)N$} scattering amplitude}
\label{app1}

Applying Feynman diagram rules for the scattering amplitude corresponding to the diagram of Fig.\ref{Fig.1}(a) one 
obtains
\begin{eqnarray}
& & \langle\lambda_{f1},\lambda_{f2},\lambda_s\mid A \mid \lambda_\gamma, \lambda_A\rangle = \nonumber \\ 
%
%N1:
& &(N1): \int {-i\Gamma_{N_{1f}}i[\sh p_{1f}-\sh k_1+m_{q}]\over 
(p_{1f}-k_1)^2-m_q^2+i\epsilon}i S(k_1)
\cdots [-igT^F_c\gamma_\mu]\cdots
{i[\sh p_{1i} - \sh k_1 + m_q](-i)\Gamma_{N_{1i}}
\over (p_{1i}-k_{1})^2-m^2+i\epsilon} {d^4k_{1}\over (2\pi)^4} \nonumber \\
%
% gamma q:
& & (\gamma q):
{i[\sh p_{1i}-\sh k_1 + q + m_q]\over (p_{1i}-k_1+q)^2-m_q^2+i\epsilon} [-iQ_ie\epsilon^\perp\gamma^\perp]
%\nonumber \\ & & \ \ \ \ \ \ \ \ \times 
\nonumber \\
%
%N2:
& &(N2): \int {-i\Gamma_{N_{2f}}i[\sh p_{2f}-\sh k_2+m_{q}]\over 
(p_{2f}-k_2)^2-m_q^2+i\epsilon}iS(k_2)\cdots 
[-igT^F_c\gamma_\nu]
{i[\sh p_{2i} - \sh k_2 + m_q](-i)\Gamma_{N_{2i}}
\over (p_{2i}-k_{2})^2-m^2+\epsilon} {d^4k_{2}\over (2\pi)^4} \nonumber \\
%
%d:
& & (^3He): \int {-i\Gamma_{^3\textnormal{\scriptsize {He}}}\cdot\bar{u}_{\lambda_s}(p_s) i[\sh p_{NN} - \sh p_{2i} + m_N]\over 
(p_{NN}-p_{2i})^2 - m_N^2+i\epsilon}
{i[\sh p_{2i}+m_N]\over p_{2i}^2-m_N^2+i\epsilon}
{d^4p_{2i}\over (2\pi)^4}\nonumber \\
%
%g:
& & (g): {i d^{\mu,\nu}\delta_{ab}\over 
[(p_{2i}-k_2)-(p_{1i}-k_1)-(q-l)]^2 + i\epsilon}, \nonumber \\
\label{amplitude}
\end{eqnarray}
where  the momenta involved above are defined in Fig.\ref{Fig.1}. 
Note that the terms above are grouped according to their momenta. As such they do not represent 
the correct sequence of the scattering presented in  Fig.\ref{Fig.1}. To indicate this we 
separated the disconnected terms  by ``$\cdots$''.

The 
covariant vertex function, $\Gamma_{^3\textnormal{\scriptsize {He}}}$ describes the transition of the $^3$He nucleus
to a three-nucleon system. The vertex function $\Gamma_N$  describes  a transition of 
a nucleon to one-quark and a residual spectator quark-gluon system with total momentum
$k_{i}$, $(i=1,2)$. The function $S(k)$ describes the propagation of the  
off-mass shell quark-gluon spectator system of the nucleon. As is shown below,
this nonperturbative function can be included in the definition of a nonperturbative 
single quark wave function of the nucleon.

Using the reference frame and the kinematic conditions described in  Sec.\ref{II} we 
now elaborate each labeled term of Eq.(\ref{amplitude}) separately.

\medskip
\medskip

\noindent {\bf ($^3$He)-term.} Using the light-cone representation of 
four-momenta and introducing the light-cone momentum fraction of 
the $NN$ pair carried by the  nucleon $2i$ as  $\alpha = {p_{2i+}\over p_{NN+}}$, one 
represents the nucleon propagators as well as the momentum integration $d^4p_{2i}$ in 
the following form:
\begin{eqnarray}
& & p_{2i}^2 - m_N^2 + i\epsilon  
= \alpha \cdot p_{NN+}(p_{2i-}- {m_N^2+p_{2i\perp}^2\over\alpha p_{NN+}})  + i\epsilon \nonumber \\
& & (p_{NN}-p_{2i})^2 - m_N^2 + i\epsilon = p_{NN+}(1-\alpha)({M_{NN}^2\over p_{NN+}} - p_{2i-}) 
- (m_N^2+p_{i\perp}^2) + i\epsilon \nonumber \\
& & d^4p_{2i} = p_{NN+}{1\over 2} d\alpha d p_{2i-} d^2p_{2i\perp}.
\label{props}
\end{eqnarray}
Using these relations in Eq.(\ref{amplitude}) we can integrate over $dp_{2i-}$ taking the residue at the pole 
of the $(2i)$-nucleon propagator, i.e.,
\begin{equation}
\int{[...]d p_{2i-}\over p_{2i-}-{m_N^2+p_{2i\perp}^2\over \alpha p_{NN+}}+i\epsilon} = 
-2\pi i[...]\mid_{p_{2i-}={m_N^2+p_{2i\perp}^2\over\alpha p_{NN+}}}.
\label{pole}
\end{equation}
After this integration one can use the following relations in Eq.(\ref{amplitude}):
\begin{eqnarray}
& & \sh p_{2i} + m_N = \sum\limits_{\lambda_{2i}} 
u_{\lambda_{2i}}(p_{2i})\bar u_{\lambda_{2i}}(p_{2i}) \nonumber \\
& &  (p_{NN}-p_{2i})^2 - m_N^2 = 
(1-\alpha)(M_{NN}^2 - {m_N^2+p_{2i\perp}^2\over \alpha(1-\alpha)})
\nonumber \\
& &  \sh p_{NN}-\sh p_{2i} + m_N = \sum\limits_{\lambda_{1i}}
 u_{\lambda_{1i}}(p_{1i})\bar u_{\lambda_{1i}}(p_{1i}) 
+ {M_{NN}^2-{m_N^2+p_{2i\perp}^2\over \alpha(1-\alpha)}\over 2p_{NN+}}\gamma^{+}.
\label{afterpole}
\end{eqnarray}
Furthermore we use the condition $p_{NN+}^2 \gg {1\over 2}(M_{NN}^2 - 
{m_N^2+p_{2i\perp}^2\over \alpha (1-\alpha)})$ to neglect the 
second term  of the right-hand part of the third equation  in Eq.(\ref{afterpole}) . 
This relation is justified for the 
high energy kinematics described in  Sec.\ref{II} as well as from the fact that 
in the discussed model the scattering amplitude is defined at $\alpha\approx {1\over 2}$.

Introducing the light-cone wave function of $^3$He \cite{FS81,FS88,polext}
\begin{equation}
\Psi_{^3\textnormal{\scriptsize {He}}}^{\lambda_A,\lambda_1,\lambda_2,\lambda_s}(\alpha,p_\perp) = {\Gamma_{^3\textnormal{\scriptsize {He}}}^{\lambda_A} 
\bar u_{\lambda_1}(p_{NN}-p)\bar u_{\lambda_2}(p)\bar u_{\lambda_s}(p_s)
\over M_{NN}^2-{m_N^2+p_\perp^2\over \alpha(1-\alpha)}}
\label{wf_d}
\end{equation}
and collecting all the terms  of Eq.(\ref{afterpole}) in the {\bf ($^3$He:)} part of 
Eq.(\ref{amplitude}) one obtains
\begin{eqnarray}
& & (^3He:) = \sum\limits_{\lambda_{1i},\lambda_{i2}} 
\int {\Psi_{^3\textnormal{\scriptsize {He}}}^{\lambda_A,\lambda_{i1},\lambda_{i2},\lambda_s}
(\alpha,p_{i\perp})\over 1-\alpha} u_{\lambda_{i1}}(p_1)u_{\lambda_{i2}}(p_2)
{d\alpha\over \alpha}{d^2p_{2i\perp}\over 2 (2\pi)^3}.
\label{^3He:}
\end{eqnarray}

\medskip
\medskip

\noindent {\bf (N1:).} To evaluate this term in Eq.(\ref{amplitude}) we first introduce
\begin{eqnarray}
x_{1} & = & {k_{1+}\over p_{1i+}} = {k_{1+}\over (1-\alpha)p_{NN+}},\nonumber \\
x_{1}^\prime & = & {k_{1+}\over p_{1f+}} = {1-\alpha\over 1-\alpha^\prime}x_1,
\label{xes}
\end{eqnarray}
where $\alpha^\prime = {p_{2f+}\over p_{NN+}}$. Furthermore we perform the $k_{1-}$ 
integration such that it puts the spectator system of the $N1$ nucleon at its
on-mass shell. This will results in
\begin{equation}
\int S(k_{1}) d k_{1-} = -{2\pi i \over p_{1+}x_1} \sum_s \psi_s(k_1)\psi^\dagger_s(k_1)
\mid_{k_{1-} = {m_s^2+k_{1\perp}^2\over p_{1+}x_1}},
\label{k1onshell}
\end{equation}
where $\psi_{s}(k)$ represents the nucleon's spectator wave function with 
mass $m_s$, and spin $s$. Note that in the definition of $\psi_{s}$ 
one assumes an integration over all the internal 
momenta of the spectator system.  Using Eq.(\ref{k1onshell}) for the (N1) term one 
obtains
\begin{eqnarray}
%N1:
(N1): & & \sum_s \int {-i\Gamma_{N_{1f}}i(\sh p_{1f}-\sh k_1+m_{q}]\over 
(p_{1f}-k_1)^2-m_q^2+i\epsilon}\psi_s(k_1)\cdots
\nonumber \\ & & 
\cdots [-igT^F_c\gamma_\mu]\psi_s^\dagger(k_1)
{i[\sh p_{1i} - \sh k_1 + m_q](-i)\Gamma_{N_{1i}}
\over (p_{1i}-k_{1})^2-m^2+i\epsilon} 
\times {dx_1\over x_1} {d^2k_{1\perp}\over 2(2\pi)^3}. 
\label{N1}
\end{eqnarray}

Now we evaluate the propagator of the off-shell 
quark with the momentum, $p_{1i}-k_{1}$. This yields:
\begin{eqnarray}
{\sh p_{1i}-\sh k_1 + m_q\over (p_{1i}-k_1)^2 - m_q^2} & = &  
{(\sh p_{1i} - \sh k_1)^{on \ shell}  + m_q \over
(1-x_1)(\tilde m_{N1}^2 - {m_s^2(1-x_{1}) + m_q^2x_1 + (k_{1\perp}-x_1p_{1\perp})^2\over
x_1(1-x_1)})}\nonumber \\
& +& {\gamma^+\over 2(1-\alpha)(1-x_{1})p_{NN+}},
\label{q1on}
\end{eqnarray}
where the effective off-shell mass of the nucleon is defined as
\begin{equation}
\tilde m_N^2 = {M_{NN}^2\alpha(1-\alpha) - m_N^2(1-\alpha)-p_{\perp}^2\over \alpha}.
\label{tilmass}
\end{equation}
As it  follows from Eq.(\ref{q1on}) at the high energy limit, $p_{NN+}^2\gg m_N^2$, one 
can neglect the second term of the RHS (off-shell) part of the equation if 
$(1-\alpha)$$(1-x_{1})$ $\sim 1$. As is shown in Sec.\ref{II} [see discussion 
before Eq.(\ref{ampl2})], the essential 
values that contribute in the scattering amplitude correspond to  $\alpha\approx {1\over 2}$ and 
$(1-x_{1})\sim 1$.  Therefore the second term in the right-hand side part of Eq.(\ref{q1on}) can be 
neglected. Using the closure relation for the on-shell spinors for Eq.(\ref{q1on}) one 
obtains
\begin{eqnarray}
{\sh p_{1i}-\sh k_1 + m_q\over (p_{1i}-k_1)^2 - m_q^2} = 
{\sum_{\eta_{1i}}u_{\eta_{1i}}(p_{1i}-k_1)\bar  u_{\eta_{1i}}(p_{1i}-k_1)
\over
(1-x_1)(\tilde m_{N1}^2 - {m_s^2(1-x_{1}) + m_q^2x_1 + (k_{1\perp}-x_1p_{1\perp})^2\over
x_1(1-x_1)})}.
\label{q1on2}
\end{eqnarray}
Similar considerations yield the following expression for the propagator of the quark entering 
the wave function of the final nucleon ``$1f$'':
\begin{eqnarray}
{\sh p_{1f}-\sh k_1 + m_q\over (p_{1f}-k_1)^2 - m_q^2} = 
{\sum_{\eta_{1f}}u_{\eta_{1f}}(p_{1f}-k_1)\bar  u_{\eta_{1f}}(p_{1f}-k_1)
\over
(1-x_1^\prime)(m_{N}^2 - {m_s^2(1-x_{1}^\prime) + m_q^2x_1^\prime + (k_{1\perp}-x_1^\prime p_{1f\perp})^2\over
x_1^\prime(1-x_1^\prime)})},
\label{q1onfin}
\end{eqnarray}
where $x_1^\prime$ is defined in Eq.(\ref{xes}). 
 
By inserting Eqs.(\ref{q1on2}) and (\ref{q1onfin}) into Eq.(\ref{N1}) and defining quark wave function of the  nucleon 
as
\begin{equation}
\Psi_N^{\lambda,\eta}(p,x,k_\perp) = {u_N^\lambda(p)\Gamma_N \bar u_\eta(p-k)\psi^\dagger_s(k)\over 
m_N^2 - {m_s^2(1-x) + m_q^2x + (k_{\perp}-xp_{\perp})^2\over
x(1-x)}}
\label{nwf}
\end{equation}
for the $(N1:)$ term we obtain
\begin{eqnarray}
(N1:) & & \sum\limits_{\eta_{1f},\eta_{1i},s_1}\int 
{\Psi^{\dagger \lambda_{1f},\eta_{1f}}(p_{1f},x_{1}^\prime,k_{1\perp}) \over (1-x_{1}^\prime)}
\bar u_{\eta_{1f}}(p_{1f}-k_1)\cdots \nonumber \\ & & 
\cdots [-igT^F_c\gamma_\mu]u_{\eta_{1i}}(p_{1i}-k_1)
{\Psi^{\lambda_{1i},\eta_{1i}} (p_{1i},x_1,k_{1\perp}) \over (1-x_1)}
{dx_1\over x_1}{d^2 k_{1\perp}\over 2 (2\pi)^3}. 
\label{N1fin}
\end{eqnarray}

\medskip
\medskip

\noindent {\bf (N2:).} This term can be evaluated following similar considerations used above in the evaluation of the (N1:) term.  Introducing light-cone momentum 
fraction of the spectator system of the second nucleon as
\begin{eqnarray}
x_{2} & = & {k_{2+}\over p_{2i+}} = {k_{2+}\over \alpha p_{NN+}},\nonumber \\
x_{2}^\prime & = & {k_{2+}\over p_{2f+}} = {\alpha\over \alpha^\prime}x_2
\label{x2es}
\end{eqnarray}
for the (N2:) term we obtain
\begin{eqnarray}
(N2:) & & \ \sum\limits_{\eta_{2f},\eta_{2i},s_2}\int 
{\Psi^{\dagger \lambda_{2f},\eta_{2f}}(p_{2f},x_{2}^\prime,k_{2\perp}) \over (1-x_{2}^\prime)}
\bar u_{\eta_{2f}}(p_{2f}-k_2)\cdots \nonumber \\
& & \cdots u_{\eta_{2i}}(p_{2i}-k_2)
{\Psi^{\lambda_{2i},\eta_{2i}} 
(p_{2i},x_2,k_{2\perp}) \over (1-x_2)}
{dx_2\over x_2}{d^2 k_{2\perp}\over 2 (2\pi)^3}.
\label{N2fin}
\end{eqnarray}

\medskip
\medskip
Collecting the expressions of Eqs.(\ref{^3He:}), (\ref{N1fin}) and (\ref{N2fin}) in Eq.(\ref{amplitude})
and rearranging terms to express the sequence of the scattering, we obtain 
the expression of the scattering amplitude presented in Eq.(\ref{ampl0}).

\section{Calculation of the nucleon-nucleon scattering amplitude}
\label{app2}

\begin{figure}[ht]
\centering\includegraphics[height=10.5cm,width=8.6cm]{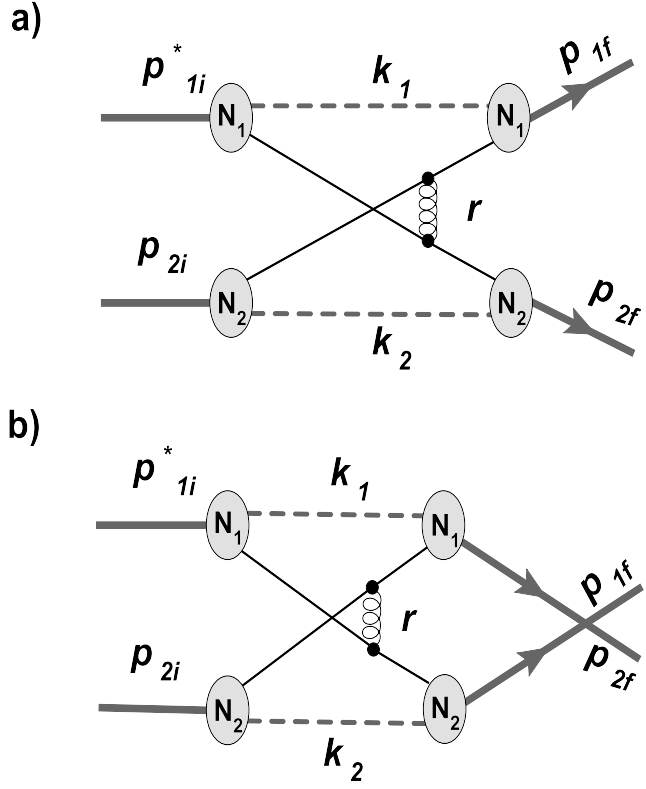}
\caption{(Color online) Quark interchange contribution to nucleon-nucleon scattering}
\label{Fig.B1}
\end{figure}

In this section we consider a hard $NN$ elastic scattering model in which two nucleons interact through 
the(QIM). 
The typical diagram for such scattering is presented in Fig. \ref{Fig.B1}.  Applying Feynman diagram rules 
for these diagrams one obtains

\begin{eqnarray}
& &  A ^{\textnormal{\scriptsize {QIM}}}_{NN} = \nonumber \\ 
%
%N1:
& &(N1): \int {-i\Gamma_{N_{1f}}i[\sh p_{1f}-\sh k_1+m_{q}]\over 
(p_{1f}-k_1)^2-m_q^2+i\epsilon}i S(k_1)
%\nonumber \\ & & \ \ \ \ \ \ \ \ \times 
\cdots [-igT^F_c\gamma_\mu] {i[\sh p^{*}_{1i} - \sh k_1 + m_q](-i)\Gamma_{N_{1i}}
\over (p^{*}_{1i}-k_{1})^2-m^2+i\epsilon} {d^4k_{1}\over (2\pi)^4} \nonumber \\
%
%N2:
& &(N2): \int {-i\Gamma_{N_{2f}}i[\sh p_{2f}-\sh k_2+m_{q}]\over 
(p_{2f}-k_2)^2-m_q^2+i\epsilon}S(k_2) \cdots
[-igT^F_c\gamma_\nu] {i[\sh p_{2i} - \sh k_2 + m_q](-i)\Gamma_{N_{2i}}
\over (p_{2i}-k_{2})^2-m^2+\epsilon} {d^4k_{2}\over (2\pi)^4} \nonumber \\
%
%g:
& & (g): {i d^{\mu,\nu}\delta_{ab}\over r^2 + i\epsilon}\nonumber \\
& & \hspace{2in} -  (p_{1f}\leftrightarrow p_{2f}),
\label{nnamp}
\end{eqnarray}
where definitions of the momenta are given in Fig.(\ref{Fig.B1}). The procedure of reducing the above 
amplitude is similar to the one used in the previous section.
First we estimate the propagators of each  nucleon's spectator system at their pole values,  
$k_{i,-}={m^{2}_{s}+k^{2}_{i,\perp}\over x_ip^+}$ ($i=1,2$) by performing the $k_{i,-}$  integration, 
which yields
\begin{eqnarray}
\int [...]S(k_i) d k_{i-} = -{2\pi i[...] \over x_ip_{+}} \sum_{s} \psi_s(k_i)\psi^\dagger_s(k_i)
\mid_{k_{i,-} = {m_s^2+k_{i,\perp}^2\over x_i p_{+}}}.
\label{kpolint}
\end{eqnarray}
Furthermore, because  $p^2_+>>m^2_N$ one can  apply similar to Eqs.(\ref{q1on2}) and (\ref{q1onfin}), approximations for 
propagators of interchanging quarks leaving and entering the corresponding nucleons. Then using the definition of 
single quark wave function according to Eq.(\ref{nwf}) for the $(N_1)$ and $(N_2)$ terms, one obtains similar 
expressions that can be presented in the following form: 
\begin{eqnarray}
(N1:) \ & & \sum\limits_{\eta_{1,2f},\eta_{2,1i},s}\int 
{\Psi^{\dagger \lambda_{1f},\eta_{1f}}(p_{1f},x_1^\prime,k_{1\perp}) \over (1-x_1^\prime)}
\bar u_{\eta_{1f}}(p_{1f}-k_1)\cdots
\nonumber \\  & & 
\cdots [-igT^F_c\gamma_{\mu}]u_{\eta_{1i}}(p^*_{1i}-k_1) 
{\Psi^{\lambda_{1i},\eta_{i1}} 
(p_{1i},x_1,k_{1\perp}) \over (1-x_1)} 
{dx_1\over x_1}{d^2 k_{1\perp}\over 2 (2\pi)^3}. 
\label{nn_N1}
\end{eqnarray}
The $(N2:)$ term is obtained from the above equation by replacing $1\rightarrow 2$.
Regrouping (N1) and (N2)  terms given by Eq.(\ref{nn_N1}) into Eq.(\ref{nnamp}),
 for the  amplitude of  nucleon-nucleon scattering in QIM we obtain
\begin{eqnarray}
& & A^{\textnormal{\scriptsize {QIM}}}_{NN} = \sum_{\eta_{1i}\eta_{2i}\eta_{1f}\eta_{2f}}\int\nonumber\\ 
& &\left[ \left\{ {\psi^{\dagger\lambda_{2f},\eta_{2f}}_N(p_{2f},x'_2,k_{2\perp})\over 1-x'_2}\bar 
u_{\eta_{2f}}(p_{2f}-k_2) [-igT^F_c\gamma^\nu]\right. 
u_{\eta_{1i}}(p^{*}_{1i}-k_1)
{\psi_N^{\lambda_{1i},\eta_{1i}}(p^{*}_{1i},x_1,k_{1\perp})\over (1-x_1)} 
\right\} \times 
\nonumber \\
& & \left\{ 
{\psi^{\dagger\lambda_{1f},\eta_{1f}}_N(p_{1f},x'_1,k_{1\perp})\over 1-x'_1}
\bar u_{\eta_{1f}}(p_{1f}-k_1)[-igT^F_c\gamma^\mu] u_{\eta_{2i}}(p_{2i}-k_2) 
 {\psi_N^{\lambda_{2i},\eta_{2i}}(p_{2i},x_2,k_{2\perp})\over (1-x_2)} \right\} \nonumber \\
& & \left. G^{\mu,\nu}(r) {dx_1\over x_1}{d^2k_{1\perp}\over 2 (2\pi)^3}
{dx_2\over x_2}{d^2k_{2\perp}\over 2 (2\pi)^3} \right].
\label{NN_qim}
\end{eqnarray}
Note that in the above expression we redefined the initial momentum of ``$N1$'' nucleon to $p_{1i}^*$ 
to emphasize its difference from $p_{1i}$ which enters in the photodisintegration amplitude. 
In the latter case $p_{1i}$ is not independent and it is defined by the momenta of the two remaining 
nucleons in the $^3$He nucleus.

\section{Calculation of Quark-Charge Factors in QIM model of $NN$ Scattering}

We consider now the hard elastic $NN$ scattering (described through $a+b\rightarrow c+d$ reaction)
within quark-interchange approximation following the approach 
used in Ref.\cite{FGST}. In this case one can represent the 
scattering amplitude as follows:
\begin{widetext}
\begin{eqnarray}
& & \langle cd\mid T\mid ab\rangle = 
\sum\limits_{\alpha,\beta,\gamma} 
\langle  \psi^\dagger_c\mid\alpha_2^\prime,\beta_1^\prime,\gamma_1^\prime\rangle
\langle  \psi^\dagger_d\mid\alpha_1^\prime,\beta_2^\prime,\gamma_2^\prime\rangle 
\nonumber \\
& & \ \ \ \  \times
\langle \alpha_2^\prime,\beta_2^\prime,\gamma_2^\prime,\alpha_1^\prime\beta_1^\prime
\gamma_1^\prime\mid H\mid
\alpha_1,\beta_1,\gamma_1,\alpha_2\beta_2\gamma_2\rangle\cdot 
\langle\alpha_1,\beta_1,\gamma_1\mid\psi_a\rangle
\langle\alpha_2,\beta_2,\gamma_2\mid\psi_b\rangle,
\label{ampl}
\end{eqnarray}
\end{widetext}
where ($\alpha_i,\alpha_i^\prime$), ($\beta_i,\beta_i^\prime$) and 
($\gamma_i\gamma_i^\prime$) describe the spin-flavor  quark states 
before and after the hard scattering, $H$,
and
\begin{equation}
C^{j}_{\alpha,\beta,\gamma} = \langle\alpha,\beta,\gamma\mid\psi_j\rangle
\label{Cs}
\end{equation}
describes the probability amplitude of finding an $\alpha,\beta,\gamma$ helicity-flavor 
combination of three valence quarks in the nucleon $j$.
The $C^{j}_{\alpha,\beta,\gamma}$ terms can be calculated by representing the nucleon wave functions
on the  helicity-flavor basis. For SU(6) symmetric wave function such representation reads
\begin{widetext}
\begin{eqnarray}
\psi^{i^3_{N},h_N} & = & {1\over \sqrt{2}}\left\{
(\chi_{0,0}^{(23)}\chi_{{1\over2},h_N}^{(1)})\cdot
(\tau_{0,0}^{(23)}\tau_{{1\over 2},i_N^{3}}^{(1)}) 
\right.  +   \nonumber \\
& & 
\sum\limits_{i_{23}^3=-1}^{1} \ \ \sum\limits_{h_{23}^3=-1}^{1}
\langle 1,h_{23}; {1\over 2},h_{N}-h_{23}\mid {1\over 2},h_N\rangle
\langle 1,i^3_{23}; {1\over 2},i^3_{N}-i^3_{23}\mid {1\over 2},i^3_N\rangle \nonumber \\
& &\left. \times (\chi_{1,h_{23}}^{(23)}\chi_{{1\over2},h_N-h_{23}}^{(1)})\cdot
(\tau_{1,i^3_{23}}^{(23)}\tau_{{1\over 2},i_N^{3}-i^3_{23}}^{(1)})\right\}
\label{wf}
\end{eqnarray}
\end{widetext}
where we separated the wave function into two parts with two quarks (e.g. 2nd and 3rd) being in helicity zero-isosinglet and helicity one-isotriplet  states.
Here $\chi_{j,h}$ is the helicity wave function with total spin $j$ and helicity $h$ and 
$\tau_{I,i^3}$ is the isospin wave function with total isospin $I$ and three-dimensional component $i^3$.
Also, $i^3_N$ and $h_N$ are  the isospin projection and helicity of the nucleon, respectively.

Assuming helicity conservation in the hard kernel of quark interactions,
\begin{equation}
H \approx \delta_{\alpha_1\alpha_1^\prime}\delta_{\alpha_1\alpha_1^\prime}
\delta_{\beta_1,\beta_1\prime}
\delta_{\gamma_1,\gamma_1^\prime}
\delta_{\beta_2,\beta_2\prime}
\delta_{\gamma_2,\gamma_2^\prime} {f(\theta)\over s^4},
\label{H}
\end{equation}
one obtains a rather simple relation for $NN$ scattering amplitude in the form\cite{FGST}
\begin{equation}
\langle cd\mid T\mid ab\rangle = Tr(M^{ac}M^{bd}),
\label{ampl2su}
\end{equation}
with
\begin{equation}
M^{i,j}_{\alpha,\alpha^\prime} = 
C^{i}_{\alpha,\beta\gamma}C^{j}_{\alpha^\prime,\beta\gamma} + 
C^{i}_{\beta\alpha,\beta}C^{j}_{\beta\alpha^\prime,\beta} + 
C^{i}_{\beta\gamma\alpha}C^{j}_{\beta\gamma\alpha^\prime},
\label{QIMMs}
\end{equation} 
where one sums over all the possible values of $\beta$ and $\gamma$.

Using the above equations for the helicity amplitudes defined in Eq.(\ref{phis}) and normalized 
according to Eq.(\ref{crs_NN}) \cite{FGST} for $pp\rightarrow pp$ one obtains
\begin{eqnarray}
\phi_1 & = & {1\over 9} (31 F(\theta,s)    +  31  F(\pi-\theta,s)) \nonumber \\
\phi_3 & = & {1\over 9} (14 F(\theta,s)    +  17  F(\pi-\theta,s)) \nonumber  \\
\phi_4 & = &-{1\over 9} (17 F(\theta,s)    +  14  F(\pi-\theta,s))
\label{pppp}
\end{eqnarray}
and for  $pn\rightarrow pn$ one obtains
\begin{eqnarray}
\phi_1 & = & {1\over 9} (14 F(\theta,s)   +  17  F(\pi-\theta,s)) \nonumber \\
\phi_3 & = & {1\over 9} (22 F(\theta,s)     +  25  F(\pi-\theta,s)) \nonumber \\
\phi_4 & = & {1\over 9} ( 8 F(\theta,s)     +   8  F(\pi-\theta,s))
\label{pnpn}
\end{eqnarray}
with $\phi_2=\phi_5=0$ due to helicity conservation.

The above-described formalism gives an explicit form for calculation of the 
charge weighted $NN$ amplitudes described in Sec.\ref{Sec.2B}.  For example 
Eq.(\ref{QNNQIM}) can be calculated by modifying Eq.(\ref{ampl2su})  
in the following form:
\begin{equation}
Q_F^{N_a}\langle cd\mid T\mid ab\rangle = \sum\limits_{\alpha,\alpha^\prime}
Q_\alpha M_{ac}(\alpha,\alpha^\prime)M_{bd}(\alpha^\prime,\alpha)),
\label{ampl2Q}
\end{equation}
where $Q_\alpha$ is the charge of the quark $\alpha $ in $|e|$ units.

\section{Helicity Amplitudes of Two-Nucleon Break-Up Reactions off $^3$He Target}

Replacing QIM amplitudes in  Eq.(\ref{ampl5}) by $NN$ helicity amplitudes of Eq.(\ref{phis}) and 
using the antisymmetry of the ground state wave function with respect to the  exchange of quantum numbers of 
any two nucleons one obtains the following expressions for the  helicity amplitudes of two nucleon 
breakup  reactions off the $^3$He nucleus,   
$\langle\lambda_{1f},\lambda_{2f},\lambda_s\mid A \mid \lambda_\gamma, \lambda_A\rangle$:\\
\noindent For a positive helicity photon,
\begin{eqnarray}
\langle+,+,\lambda_s\mid A \mid +, \lambda_A\rangle & = & B \int\left[Q_F\phi_5 \Psi_{^3\textnormal{\scriptsize {He}}}^{\lambda_A}(+,-,\lambda_s) + 
Q_F\phi_1\Psi_{^3\textnormal{\scriptsize {He}}}^{\lambda_A}(+,+,\lambda_s)\right]m_N {d^{2}p_{\perp}\over (2\pi)^2}\nonumber \\
\langle+,-,\lambda_s\mid A \mid +, \lambda_A\rangle & = & B  \int
\left[(Q_F^{N_1}\phi_3 + Q_F^{N_2}\phi_4) \Psi_{^3\textnormal{\scriptsize {He}}}^{\lambda_A}(+,-,\lambda_s) - 
Q_F\phi_5\Psi_{^3\textnormal{\scriptsize {He}}}^{\lambda_A}(+,+,\lambda_s)\right]\nonumber \\ 
& &  \hspace{3.6in}  \times m_N {d^{2}p_{\perp}\over (2\pi)^2}\nonumber \\
\langle-,+,\lambda_s\mid A \mid +, \lambda_A\rangle & = & B  \int
\left[-(Q_F^{N_1}\phi_4 + Q_F^{N_2}\phi_3) \Psi_{^3\textnormal{\scriptsize {He}}}^{\lambda_A}(+,-,\lambda_s) +
Q_F\phi_5\Psi_{^3\textnormal{\scriptsize {He}}}^{\lambda_A}(+,+,\lambda_s)\right]\nonumber \\
& & \hspace{3.6in } \times m_N {d^{2}p_{\perp}\over (2\pi)^2}\nonumber \\
\langle-,-,\lambda_s\mid A \mid +, \lambda_A\rangle & = & B \int\left[Q_F\phi_5 \Psi_{^3\textnormal{\scriptsize {He}}}^{\lambda_A}(+,-,\lambda_s) + 
Q_F\phi_2\Psi_{^3\textnormal{\scriptsize {He}}}^{\lambda_A}(+,+,\lambda_s)\right]m_N {d^{2}p_{\perp}\over (2\pi)^2}\nonumber \\
\end{eqnarray}
and for a negative helicity photon,
\begin{eqnarray}
\langle+,+,\lambda_s\mid A \mid -, \lambda_A\rangle & = & -B \int\left[-Q_F\phi_5 \Psi_{^3\textnormal{\scriptsize {He}}}^{\lambda_A}(-,+,\lambda_s) + 
Q_F\phi_2\Psi_{^3\textnormal{\scriptsize {He}}}^{\lambda_A}(-,-,\lambda_s)\right]m_N {d^{2}p_{\perp}\over (2\pi)^2}\nonumber \\
\langle+,-,\lambda_s\mid A \mid -, \lambda_A\rangle & = & -B  \int
\left[-(Q_F^{N_1}\phi_4 + Q_F^{N_2}\phi_3) \Psi_{^3\textnormal{\scriptsize {He}}}^{\lambda_A}(-,+,\lambda_s) - 
Q_F\phi_5\Psi_{^3\textnormal{\scriptsize {He}}}^{\lambda_A}(-,-,\lambda_s)\right] \nonumber \\ 
& & \hspace{3.8in} \times m_N {d^{2}p_{\perp}\over (2\pi)^2}\nonumber \\
\langle-,+,\lambda_s\mid A \mid -, \lambda_A\rangle & = & -B  \int
\left[(Q_F^{N_1}\phi_3 + Q_F^{N_2}\phi_4) \Psi_{^3\textnormal{\scriptsize {He}}}^{\lambda_A}(-,+,\lambda_s) +
Q_F\phi_5\Psi_{^3\textnormal{\scriptsize {He}}}^{\lambda_A}(-,-,\lambda_s)\right]\nonumber \\
& & \hspace{3.8in} \times m_N {d^{2}p_{\perp}\over (2\pi)^2}\nonumber \\
\langle-,-,\lambda_s\mid A \mid -, \lambda_A\rangle & = & -B \int\left[-Q_F\phi_5 \Psi_{^3\textnormal{\scriptsize {He}}}^{\lambda_A}(-,+,\lambda_s) + 
Q_F\phi_1\Psi_{^3\textnormal{\scriptsize {He}}}^{\lambda_A}(-,-,\lambda_s)\right]m_N {d^{2}p_{\perp}\over (2\pi)^2},\nonumber \\
\end{eqnarray}
where $B = {ie\sqrt{2}(2\pi)^3\over \sqrt{2s^\prime_{NN}}}$.
Because the scattering process is considered in the   ``$\gamma$-$NN$'' center of mass  reference frame where 
$z$ direction is chosen opposite to the momentum of incoming photon,  the bound nucleon helicity states correspond to 
the nucleon spin projections ${1\over 2}$ for positive  and $-{1\over 2}$ for negative helicities.

\end{document}